\documentclass[twocolumn,astrosymb]{aastex7}

\usepackage{amsmath}
\usepackage{gensymb}

\usepackage{subfigure}

\begin{document}

\title{Constructing Earth Formation History Using Deep Mantle Noble Gas Reservoirs}

\author[orcid=0000-0002-2870-6940,gname=Vincent,sname=Savignac]{Vincent Savignac}
\affiliation{Department of Astronomy \& Astrophysics, UC San Diego, La Jolla, CA, 92093-0424, USA}
\affiliation{Department of Physics, McGill University, 3600 rue University, Montreal, QC, H3A 2T8, Canada}
\email[show]{vsavignac@ucsd.edu}  

\author[orcid=0000-0002-1228-9820,gname=Eve,sname=J. Lee]{Eve J. Lee}
\affiliation{Department of Astronomy \& Astrophysics, UC San Diego, La Jolla, CA, 92093-0424, USA}
\affiliation{Department of Physics, McGill University, 3600 rue University, Montreal, QC, H3A 2T8, Canada}
\email{evelee@ucsd.edu}

\begin{abstract}
    Noble gases are powerful probes of the Earth’s early history, as they are chemically inert. Neon isotopic ratios in deep mantle plumes suggest that nebular gases were incorporated into the Earth’s interior. This evidence implies the Earth’s formation began when there was still gas around, with Earth embryos accreting primordial gas and a fraction of that gas dissolved into molten magma. 
    In this work, we examine these implications, simulating the growth of primordial envelopes using modern gas accretion schemes, and computing the dissolution of nebular Ne into magma oceans following chemical equilibrium.
    We find that the embryo mass that reproduces the deep mantle concentration of primordial Ne is tightly constrained to $\sim 0.3-0.4M_\oplus$, within a solar nebula depleted by $\geq10\times$ in gas density. Embryos of smaller masses cannot accrete enough gas to allow the mantle to reach the melting temperature of basalt. Embryos of larger masses accrete way too much gas, producing excessive Ne concentrations in the deep mantle.
    Based on our calculations, we suggest that the Earth’s formation began with the assembly of $\sim 0.3-0.4M_\oplus$ embryos during the dispersal of the solar nebula. 
    Light noble gases (He, Ne) in the deep mantle reflect the primordial gas accretion history of the Earth, while heavy noble gases (Ar, Kr, Xe) probe early solid accretion processes.
    Our results are consistent with the final assembly of the Earth through at least two giant impacts after the dispersal of the nebula.
\end{abstract}



\section{Introduction} 
\label{sec:introduction}

Noble gases are direct geochemical probes of the Earth's early accretion processes because of their limited interactions with other elements. Their chemical inertness allows them to be transported through planetary interiors without altering the gas composition, making them ideal tracers of volatile delivery. 
Noble gas reservoirs can be found in different layers of the inner Earth \citep[e.g.,][]{Marty_2012,Halliday_2013}, with trace amounts detectable at the surface (atmosphere plus continental crust), in mid-ocean ridge basalts (MORB; degassed upper mantle), and in ocean island basalts (OIB; deep mantle plumes). 
Furthermore, non-radiogenic isotopes of noble gases (e.g., $^3$He,  $^{20}$Ne, $^{36}$Ar, $^{84}$Kr, and $^{130}$Xe) trapped in MORB and OIB samples are not produced by radioactive decay and do not themselves undergo decay on timescales relevant to planetary evolution, which implies their existence since the formation of the solar system \citep[e.g.,][]{Trieloff_2005}.
Measuring the isotopic ratios of non-radiogenic species found in MORB and OIB samples is therefore a powerful tool for assessing the origin of these gases in the Earth's mantle and the accretion history of the planet.

However, the origin of the deep mantle noble gas reservoirs remains debated \citep[see, e.g.,][for a review]{Day_2022}, with three main possible explanations for their abundances. First, the nebular capture hypothesis suggests that the formation of Earth embryos in the early presence of the solar nebula would have led to the capture of primordial gas \citep{Mizuno_1980,Abe_1985,Sasaki_1990}. A proto-atmosphere, accreted early in the primordial solar nebula, could have dissolved a fraction of its gas into the hot, molten surface of planetary embryos. Second, solar-wind-irradiated (SWI) materials, that is, solids enriched with solar-wind ions, were incorporated into the Earth in the early stages of the solar system \citep{Trieloff_2000,Ballentine_2005,Raquin_2009,Colin_2015,Moreira_2016,Peron_2016,Peron_2018}. In fact, \cite{Peron_2017} showed by comparing samples of irradiated lunar soil with MORB samples that the upper mantle reservoirs of Ne, He and possibly H are expected to be of SWI origin. Finally, carbonaceous chondrites, which are primitive meteorites rich in volatile elements \citep{Alaerts_1979a,Alaerts_1979b,Kallemeyn_1981,Wieler_1991,Wieler_1992,Huss_1996,Busemann_2000}, may also explain the noble gas reservoirs of the deep mantle, assuming they were incorporated into the Earth's interior during its assembly \citep{Holland_2009,Marty_2012,Halliday_2013}.

For an imprint of these initial reservoirs to remain measurable in the mantle, all three scenarios require limited interactions between the mantle and the secondary atmosphere. However, the subduction of atmospheric material from the oceanic lithosphere to the mantle is an efficient carrier of the heavier noble gases \citep{Peron_2022}.
As a result, the Kr and Xe mantle budget suffers from significant atmospheric contamination, limiting studies of the origins of the initial Kr and Xe reservoirs \citep{Ballentine_2000,Mukhopadhyay_2019}.
Lighter noble gases such as He and Ne do not show traces of subducted atmospheric material \citep{KENDRICK_2018}, as their poor solubility from the atmosphere into seawater \citep{Sander_2015} hinders their transfer from the oceanic lithosphere to the interior. Consequently, He and Ne offer the most promising avenue for unveiling the imprint of early accretion processes in the deep mantle.

Measurements of the stable isotopes of Ne ($^{20}$Ne, $^{21}$Ne and $^{22}$Ne) are consistent with a solar-like pristine neon reservoir preserved in the deep mantle. High isotopic ratios ${\rm ^{20}Ne/^{22}Ne }> 12$ rule out atmospheric contamination or chondrites as the predominant source, with the expected ${\rm ^{20}Ne/^{22}Ne }$ values of $9.80 \pm 0.01$ and $10.67 \pm 0.02$, respectively \citep[from Table 1 of ][]{Marty_2022}. Whether the deep mantle signature corresponds to remnants of the solar nebula or SWI material is still debated, with the former scenario characterized by ${\rm ^{20}Ne/^{22}Ne }=13.36 \pm 0.18$ \citep{Heber_2012} and the latter $12.7 \pm 0.1$ \citep{Moreira_2016}.

On one side, analysis of OIB samples have constrained the ${\rm ^{20}Ne/^{22}Ne }$ signature of  Galapagos plume mantle sources to $12.65 \pm 0.04$ \citep{Peron_2017} and more recently that of Iceland to $12.63 \pm 0.14$ \citep{Sauvalle_2026}, which are in direct agreement with the implanted solar wind model. 
Meanwhile, the step-crushing analysis of \cite{Williams_2019} of samples of the Discovery plume in the south Atlantic yields high ${\rm ^{20}Ne/^{22}Ne }$ ratios of $12.83 \pm 0.05 \ (2\sigma)$ and $13.03 \pm 0.04 \  (2\sigma)$, which exceed the SWI prediction and thus require a nebular component. In their comparison of different plume samples with the depleted MORB mantle using a two-component mixing array, \cite{Williams_2019} concluded that the deep mantle source of Ne must be a mixture of a small chondrite-like component and a pristine source with a $^{20}$Ne/$^{22}$Ne isotopic ratio of $13.23 \pm 0.22$ (see their figure 3). This ratio is statistically indistinguishable from the nebular capture expectation and deviates by more than 2$\sigma$ from the SWI scenario. Such a finding is also in agreement with the previous lower bound of ${\rm ^{20}Ne/^{22}Ne}\geq13.0 \pm 0.2$ for pristine neon in the deep mantle established by \cite{Yokochi_2004} and offers a strong argument for a Ne reservoir in the deep mantle dominated by primordial nebular gas. As a counter argument, \cite{Sauvalle_2026} suggested that the SWI value ${\rm ^{20}Ne/^{22}Ne }=12.7 \pm 0.1$ predicted by \cite{Moreira_2016} is likely to vary with the exposure time of Earth's parent bodies to solar wind and may still explain higher ratios of ${\rm ^{20}Ne/^{22}Ne} \approx 13$. We note that this debate remains unsettled and that the two supply mechanisms of pristine Ne to the deep mantle are not mutually exclusive.

A reservoir of nebular gas in the interior of the Earth would have strong implications for its formation history.
The first one is that Earth embryos must have emerged when the gaseous solar nebula was still around, suggesting that the formation of the Earth happened early with the accretion of primordial atmospheres.
Furthermore, the capture of volatiles in the planetary interior necessitates an exchange of gas between gas envelopes and planetary mantles via dissolution, requiring the presence of liquid magma oceans at the mantle-envelope boundary. Planetary embryos must therefore have been sufficiently massive for the accretion of an atmosphere and for their rocky interior to have a molten outer layer.
The cooling of the accreted gas sets the thermal conditions at the base of the atmosphere, which in turn control how much of the atmospheric volatiles are incorporated into the mantle through the atmosphere-mantle interface.
For that reason, realistic envelope formation models in the nebula are required to accurately link the formation environment of protoplanets to their interior reservoirs of noble gases.

In this work, we adopt the conclusion of \cite{Williams_2019} and assess whether there exists a self-consistent physical explanation for nebular Ne signatures in the deep mantle, the formation history of the Earth and the origin of other volatile reservoirs. 
We first construct in Section \ref{sec:methods} an envelope formation model of Earth embryos embedded in the protoplanetary disk of the Sun based on modern gas accretion models. We then connect the thermal state of forming primordial gas envelopes to the corresponding concentration of nebular gas dissolved in the mantle that would be measurable in deep mantle plumes. Section \ref{sec:results} presents our results, placing constraints on the properties of early Earth embryos, and assesses the consistency of our calculation with the mixing and cooling of the silicate mantle of young embryos.
In Section \ref{sec:formation}, we discuss whether our results can be reconciled with the current understanding of the formation of the Earth, from giant impacts after the nebular gas dissipates. We discuss how our results can be reconciled with the measured interior reservoirs of other noble gases in Section \ref{sec:Noble_gas_budget} and predict in Section \ref{sec:H_dissolution} the concentration of primordial hydrogen delivered to the Earth's interior.
Finally, we summarize our findings in Section \ref{sec:conclusion}.

\section{Methods} \label{sec:methods}

\begin{figure}[t]
\centering
\includegraphics[width=8cm]{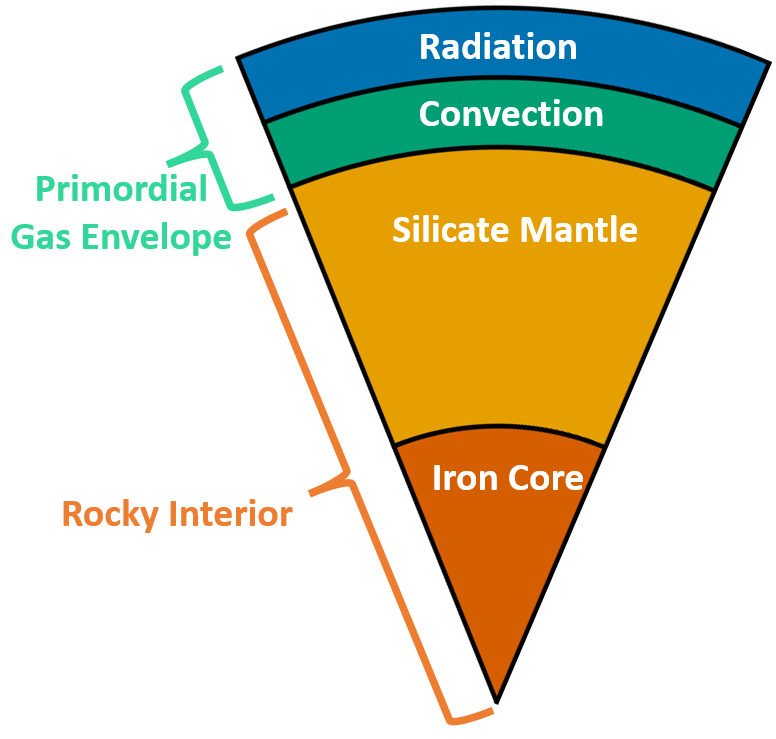}
\caption{Layered structure of Earth embryos considered in this work. Embryos are divided into a rocky interior and a surrounding gas envelope accreted from the primordial nebula of the solar system. We assume that the rocky interior is made of an innermost iron core and an outer silicate mantle, analogous to the Earth's current internal structure. As argued in Section \ref{subsec:gas_accretion}, the energy transport within the inner and outer atmospheric layers is set by convection and radiation, respectively, which follows from the Schwarzschild criterion. Note that the figure is not to scale. In reality, the envelope dominates the volume despite its small contribution to the total embryo mass.
\label{fig:structure_diagram}}
\end{figure}

In this section, we build a formation model of Earth embryos, self-consistently connecting the primordial solar system nebula to the observed nebular signatures of $^{22}$Ne in the Earth's deep mantle.\footnote{ $^{22}$Ne is effectively non-radiogenic in the mantle, as total nucleogenic production \citep[$\sim 1.3 \times 10^{-18} {\rm mol/g}$ according to][]{Yatsevich_1997} is negligible relative to the bulk mantle inventory \citep[$\sim 5 \times 10^{-15} {\rm mol/g}$ according to][]{Marty_2012}.} 
We use the following nomenclature to describe the internal structure of Earth embryos, as depicted in Figure \ref{fig:structure_diagram}. The rocky interior consists of a silicate mantle that overlays an iron core, and the mantle + core structure accretes a surrounding gaseous envelope for which we solve with structure equations.
As shown in Section \ref{subsec:gas_accretion}, our calculations yield an atmosphere divided into a convective interior and a radiative exterior.\footnote{Hydrodynamical studies \citep{Lambrechts_2017,Popovas_2018,Bethune_2019,Moldenhauer_2021,Bailey_2023} of envelope-disk interactions report an additional advective outer layer in the super-Earths/sub-Neptunes regime, within which recycling flows from the disk restrict the envelope to $\sim 30\%$ of its radius. Repeating the advection calculations of \cite{Savignac_2024}, we find that atmospheric recycling can be ignored in this study of Earth embryos, since the additional advective layers rapidly cool, rendering them effectively indistinguishable from the otherwise near-isothermal radiative outer envelope. \label{footnote:advection}}

We first develop in Section \ref{subsec:gas_accretion} a gas accretion calculation of primordial atmospheres atop Mars to Earth-sized rocky protocores embedded in the primordial nebula. 
We subsequently compute in Section \ref{subsec:dissolution} the dissolved concentration of nebular Ne in the silicate mantle, regulated by the thermal state of the mantle-envelope boundary. 
Only a selection of runs for rocky interior masses $\leq 0.4M_\oplus$ are presented in Section \ref{sec:results}, as this subset of our simulations are the one yielding relevant results to the problem at hand.

Our calculations work under the assumptions that Earth embryos remain at fixed orbital distance during gas accretion and that dust grains do not contribute to nebular opacities. Both assumptions are valid in a late-stage gas-depleted (but not gas-empty) environment, which we confirm to be a preferred solution a posteriori, see Section \ref{subsubsec:late_stage_formation}.
Potential opacity contributions from dust grains are discussed in Section \ref{subsubsec:gas_rich_formation}.
Lastly, our model is constructed such that embryos with equal interior volatile concentrations coalesce, preserving their composition from the initial to the final stage of the Earth's assembly (see Section \ref{subsec:Earth_final_assembly}).

\subsection{Gas accretion model}
\label{subsec:gas_accretion}

Based on the widely accepted bottom-up core accretion scenario of \citet{Pollack_1996}, we implement one-dimensional quasi-hydrostatic gas accretion calculations to track the envelope mass growth of planetary embryos embedded in primordial disks of gas and dust. We adapt the model of \cite{Lee_2014}, initially built to characterize the gas accretion processes of super-Earths, to study smaller, gas-poor Earth embryos. 
As discussed in Footnote \ref{footnote:advection}, adding an additional upper advective layer \citep{Savignac_2024} makes no difference in this problem.
Here, we summarize the key features of the calculation and highlight the adjustments made in this study; for complete details, readers are referred to \cite{Lee_2014}.

The process consists of building a series of hydrostatic ``snapshots" of a primordial gas envelope of mass $M_{\rm env}$, accreted onto a rocky interior of mass $M_{\rm rock}$. 
Each of these snapshots is obtained by integrating the standard stellar structure equations,

\begin{equation}
    \label{eq:dM}
    \frac{dM(<r)}{dr}=4\pi r^2 \rho,
\end{equation}

\begin{equation}
    \label{eq:dP}
    \frac{dP}{dr}=-\frac{GM(<r)}{r^2}\rho -\frac{GM_{\odot}r}{a^3}\rho,
\end{equation}

\begin{equation}
    \label{eq:dT}
    \frac{dT}{dr}=\frac{T}{P} \frac{dP}{dr} \nabla,
\end{equation}
for the pressure $P$, density $\rho$, temperature $T$ and enclosed mass $M(<r) \equiv M$ of the gas envelope plus the rocky interior as functions of the planetary radius $r$. Here, $G$ is the gravitational constant, $a$ is the fixed orbital distance and $\nabla \equiv d \ln T/d \ln P$ is the dimensionless temperature gradient of the primordial gas. 
As suggested by \cite{Zhu_2021}, we account for the effect of the Sun's gravitational field on the hydrostatic balance of the envelope. The ensuing additional term in Equation (\ref{eq:dP}) is effectively a minor correction, especially at $a > 0.1$ au around a solar mass star.
 
We assume the nebular composition to be solar \citep{Grevesse_1993}, with fixed values of $X=0.70$, $Y=0.28$ and $Z=0.02$ for the relative mass fractions of H, He and metals (i.e. elements heavier than H and He), respectively.
Within the atmosphere, we implement the ideal equation of state (EOS) of \cite{Lee_2014}, which accounts for the different molecular, atomic, and ionized states of H as functions of temperature and pressure, as well as atomic He and metallic species at the aforementioned abundance rate.

The rate at which a forming planet accretes gas from the surrounding nebula is controlled by the thermal contraction of its gas envelope \citep[e.g.,][]{Pollack_1996,Ikoma_2000,Lee_2014,Piso_2015}. We thus carefully make use of a realistic treatment of the cooling of the envelope with the gradient $\nabla$ set by the dominant form of energy transport, either radiation 
\begin{equation}
    \label{eq:grad_rad}
    \nabla_{\rm rad} = \frac{3\kappa P}{64 \pi GM \sigma_{\rm SB} T^4}L,
\end{equation}
or convection 
\begin{equation}
    \label{eq:grad_ad}
    \nabla_{\rm ad} = - \frac{\partial \mathrm{log} S}{\partial \mathrm{log}P}
    \Big |_T \Big (\frac{\partial \mathrm{log} S}{\partial \mathrm{log}T}
    \Big |_P \Big ) ^{-1}.
\end{equation}  
Here, $\kappa$ is opacity, $\sigma_{\rm SB}$ is the Stefan-Boltzmann constant, $L$ is the luminosity of the envelope, and $S$ is the specific entropy of the gas.
We set $\nabla = \nabla_{\rm ad}$ when the Schwarzschild criterion is satisfied
\begin{equation}
\label{eq:schwarchsild}
    \nabla_{\rm rad} > \nabla_{\rm ad},
\end{equation}
assuming no compositional gradients, and $\nabla = \nabla_{\rm rad}$ otherwise.

The opacity of the nebular gas has a significant effect on its cooling, and by extension gas accretion rates \citep{Ikoma_2000,Lee_2014,Piso_2015}. In late-stage protoplanetary disks such as the one in which primordial Earth embryos are presumed to accrete their atmosphere (see Section \ref{subsubsec:late_stage_formation}), dust grains are expected to have coagulated and rained out so that they no longer contribute to the opacity \citep{Mordasini_2014,Ormel_2014}. We therefore adopt a ``dust-free" opacity model based on the tabulated values of \cite{Ferguson_2005} and \cite{Freedman_2014}. In the absence of dust grains, molecular opacities dominate at low temperatures below the H$_2$ dissociation front ($T \sim 2500 \mathrm{K}$), and H$^-$ opacities otherwise dominate the higher temperature regimes. 
Section \ref{subsubsec:gas_rich_formation} discusses the impact of potential opacity contributions for dust grains.

A major simplification of our calculation is the assumption that the envelope luminosity $L$ is spatially invariant for each hydrostatic profile, which implicitly neglects the contribution of the outer radiative layers.
This choice is justified by the centrally concentrated mass structure of the envelope due to H$_2$ dissociation driving the adiabatic index $\gamma < 4/3$, localizing the thermal inertia of the atmosphere to its inner convective layers. 
Following \cite{Piso_2014,Lee_2014}, we verify a posteriori the validity of this assumption by computing the neglected luminosity generated in the outer radiative envelope,

\begin{equation}
\label{eq:L_negl}
    L_{\rm negl} = -\int_{\rm rad} \rho T \frac{\Delta S_M}{\Delta t}  4 \pi r^2 dr.
\end{equation}
Here $\Delta S_M$ is the difference of $S$ evaluated at a given mass $M$ between snapshots separated by a time $\Delta t$. We find that for dust-free envelopes at $\sim 1$ au, we can safely ignore $L_{\rm negl}$ since $L_{\rm negl}/L \lesssim 10\%$.

Consequently, each hydrostatic snapshot can be assigned a constant luminosity eigenvalue $L$ that is solved for iteratively until the mass profile agrees with the fixed value of $M_{\rm env}$ within a relative difference of 1\%.
We construct snapshots for an increasing series of $M_{\rm env}$, which are then threaded in time by calculating the cooling time from one snapshot to the next. 
We ignore the thermal energy released from the rocky interior and return to the implications of this assumption in Section \ref{subsec:limations}.

For the boundary condition, we adopt the fiducial minimum-mass solar nebula (MMSN) parameters for the gas density and temperature profiles of \cite{Hayashi_1981},

\begin{equation}
    \label{eq:rho_MMSN}
    \rho_{\rm disk} = 1.4 \times 10^{-6} \mathrm{\ kg }\ \mathrm{ m}^{-3} \Big(\frac{a}{1 \mathrm{au}} \Big)^{-2.75} f_{\rm dep},
\end{equation}

\begin{equation}
    \label{eq:T_MMSN}
    T_{\rm disk} = 280 \mathrm{K} \Big(\frac{a}{1 \mathrm{au}} \Big)^{-1/2},
\end{equation}
evaluated at a fixed orbital separation of $a = 1$ au, 
where
$f_{\rm dep} \leq 1$ is a gas depletion factor we vary, representing different evolutionary stages of nebular disks.
We leave $f_{\rm dep}$ as a free parameter, constrain its possible values in Section \ref{sec:results}, and discuss its relation to formation timescales in Section \ref{subsec:embryo_formation}. 

The state variables $(P,T,\rho)$ smoothly transition from the outer nebular conditions to the envelope, and the differential system is integrated inward to the surface of the silicate mantle, at $R_{\rm rock} \equiv R_\oplus \Big( \frac{M_{\rm rock}}{M_\oplus } \Big)^{1/3}$ \citep{VALENCIA2006545}.
The outer boundary of the envelope is set by the Bondi radius, as appropriate for low mass sub-Earth embryos at 1 au

 \begin{equation}
    \label{eq:R_B}
    R_B = \frac{GM_p}{c_s^2} \simeq 64 R_\oplus \Big( \frac{M_p}{M_\oplus} \Big) \Big( \frac{\mu_{\rm disk}}{2.37} \Big)\Big(\frac{a}{1\ \mathrm{au}} \Big)^{1/2},
\end{equation}
where $M_p \equiv M_{\rm rock} +M_{\rm env}$ is the total planetary mass, the sound speed 
 \begin{equation}
    \label{eq:c_s}
    c_s = \sqrt{\frac{k_B T_{\rm disk}}{\mu_{\rm disk} m_{\rm H}}}
\end{equation}
is set by the disk temperature (Equation \ref{eq:T_MMSN}), the mean molecular weight of the MMSN disk ($\mu_{\rm disk} \approx 2.37$) is computed from the EOS, $k_B$ is the Boltzmann constant and $m_{\rm H}$ the mass of the hydrogen atom.

\begin{figure}[h]
\subfigure{\includegraphics[width=8cm]{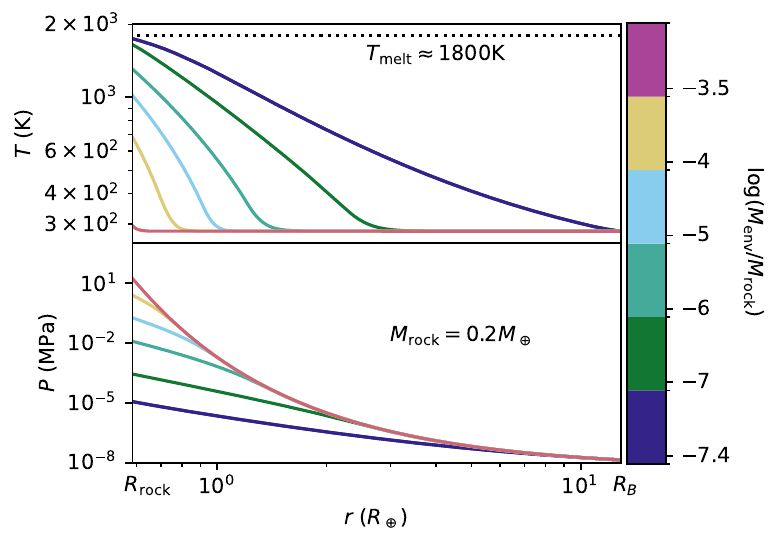}}
\subfigure{\includegraphics[width=8cm]{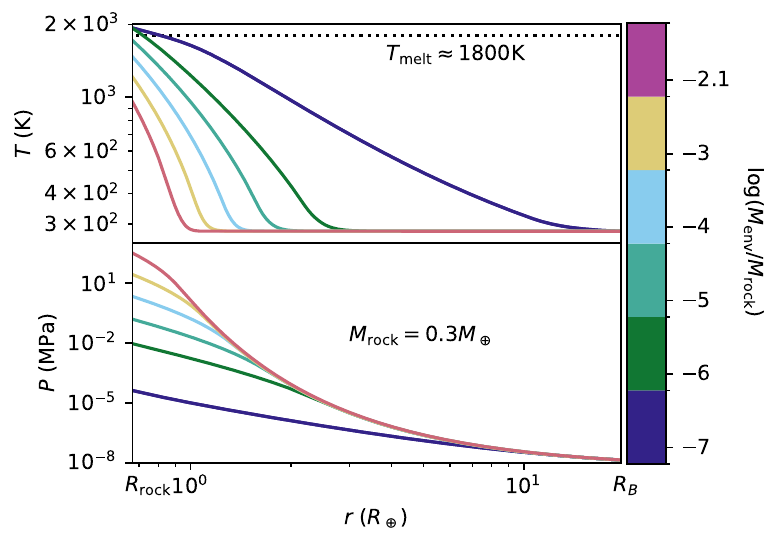}}
\subfigure{\includegraphics[width=7.3cm]{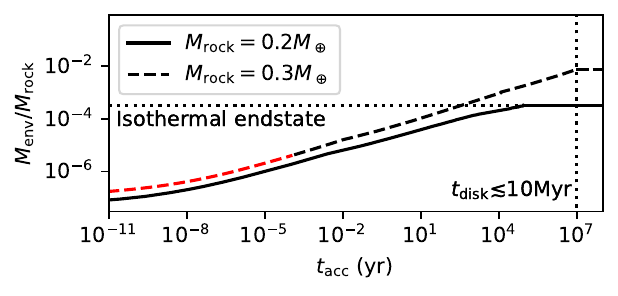}}

\caption{Formation of primordial gas envelopes atop rocky interiors embedded at 1 au in the minimum-mass solar nebula of \cite{Hayashi_1981}, with gas density depleted by a factor of $f_{\rm dep} = 10^{-2}$. 
Top: Radial profiles (solid) of the temperature and pressure $(T,P)$ of the envelope with mass $M_{\rm env}$ for a rocky interior of mass $M_{\rm rock}=0.2M_\oplus$, extending from the radius of the rocky interior $R_{\rm rock}$ to the Bondi radius $R_B$. 
Each color represents a hydrostatic snapshot of the envelope for specific values of $M_{\rm env}/M_{\rm rock}$. The dotted line indicates the melting temperature $T_{\rm melt} \approx 1800 \mathrm{K}$  of the basaltic mantle. 
Center: Same as top, but for $M_{\rm rock}=0.3M_\oplus$. Bottom: Envelope-to-rock mass ratio $M_{\rm env}/M_{\rm rock}$ over the accretion timescale $t_{\rm acc}$ for $M_{\rm rock}=0.2M_\oplus$ (solid) and $M_{\rm rock}=0.3M_\oplus$ (dashed). A dotted vertical line shows the maximal disk lifetime \citep{Mamajek_2009} and a horizontal dotted line denotes the isothermal endstate of the $M_{\rm rock}=0.2M_\oplus$ case. Values of $t_{\rm acc}$ for which the rocky interior is molten are indicated in red.}

\label{fig:example_envelope}
\end{figure}

Figure \ref{fig:example_envelope} shows examples of the envelope formation calculation for $M_{\rm rock} = 0.2 M_\oplus$ and $M_{\rm rock} = 0.3 M_\oplus$ Earth embryos accreting gas from a depleted nebula with $f_{\rm dep} = 10^{-2}$. In both cases, the initially fully convective envelope cools, releasing its thermal energy in the disk as radiative near-isothermal layers begin to grow in the outer regions. Accretion is halted either when the envelope cools down to a near-isothermal end state (e.g., top panel of Figure \ref{fig:example_envelope}) or when the gas disk with typical lifetimes of $\lesssim 10 \mathrm{Myr}$ \citep{Mamajek_2009,Alexander_2014,Michel_2021}  dissipates (e.g., middle panel of Figure \ref{fig:example_envelope}).

\subsection{Dissolution of primordial gas in the deep mantle}
\label{subsec:dissolution}

We develop a dissolution calculation under the assumption that a fraction of the total Ne budget of the primordial envelope is stored in the mantle during accretion, following equilibrium chemistry (i.e., balance of chemical potentials between dissolved and undissolved gas).
From Henry's law, the primordial concentration $c_{\rm ^{22}Ne,p}$ of nebular $^{22}$Ne initially dissolved in the melt is 

\begin{equation}
    \label{eq:Henrys_law}
    c_{\rm ^{22}Ne,p} = \frac{n_{\rm ^{22}Ne} P_0}{\mu^{M_{\rm Basalt}}k_{\rm Ne}}.
\end{equation}
Here, $n_{\rm ^{22}Ne}$ is the elemental abundance of $^{22}$Ne in the solar nebula, $P_0 \equiv P(R_{\rm rock})$ is the surface pressure of the envelope-mantle boundary, $\mu^{M_{\rm Basalt}}$ is the molar mass of the molten silicate mantle and $k_{\rm Ne}$ is the Henry's Law constant of Ne in basaltic melts.
We assume that different isotopes of an element share the same solubility constant, in the absence of evidence to the contrary.

Atmospheric volatiles dissolved at the surface of magma oceans must be able to fully mix through the rocky interior to explain primordial signatures of the Earth's deep mantle. Therefore, not only must magma oceans exist at the envelope-mantle boundary during the mixing stage, but the entire silicate mantle itself must be in a molten state to rapidly circulate volatiles to the deep interior via convection. 
It is reasonable to assume that Mars-sized embryos emerge with $T_0 \gtrsim 2000 {\rm K}$, considering the $\gtrsim 1600{\rm K}$ temperatures at which smaller $\sim 100-500 {\rm \ km}$ planetesimals are expected to reach over their early accretion \citep{Ricard_2017,Kaminski_2020,Dodds_2021}. 
In fact, \cite{Ricard_2017} conclude from thermal evolution simulations of forming planetesimals that emerging proto-planets reach temperatures of $\gtrsim 2000{\rm K}$ as long as their assembly is completed within $\lesssim 4 {\rm Myr}$ (see their Figures 10 and 12). By construction, the early Earth embryos considered in this work must coexist with the nebula prior to its dispersal (which takes $\leq$10 Myr), and can therefore be assumed to emerge with $T_0 \gtrsim 2000 {\rm \ K}$.

As the initially hot mantle of an embryo cools, the increase in pressure leads to a solidification front emerging at low depths and progressively growing outward to the surface. As such, envelope-mantle interactions are no longer a viable delivery method of nebular Ne to the bulk mantle once the interior thermal profile crosses silicate melting curves.

\begin{figure}[t]
\subfigure{\includegraphics[width=7.5cm]{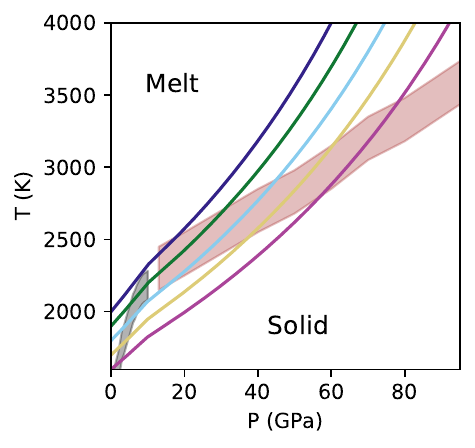}}
\subfigure{\includegraphics[width=7.5cm]{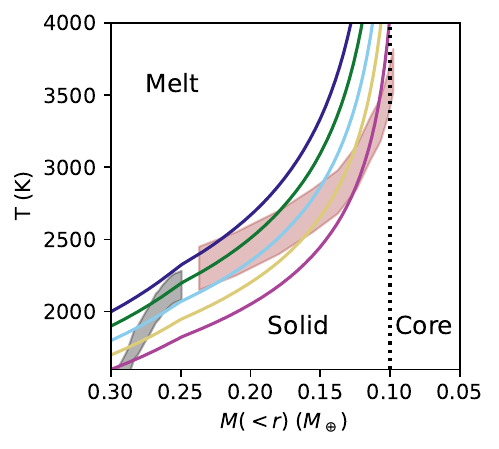}}

\caption{ Adiabatic thermal profiles of the silicate interior of $M_{\rm rock} = 0.3 M_\oplus$ Earth embryos compared to mantle solidus curves. Upper panel: Temperature $T$ versus pressure $P$ profiles (solid curves) for surface temperatures of $[1600, 1700, 1800, 1900, 2000]$K in magenta, yellow, cyan, green and blue, respectively. The profiles are compared with constraints on the solidus point of peridotite for $P\leq 10 {\rm GPa}$ \citep{Hirschmann_2000} and of a chondritic Earth-like deep mantle for $P> 12 {\rm GPa}$ \citep{Andrault_2011} displayed with shaded black and red areas, respectively. Lower panel: Same as the upper panel, but the temperature profiles are now shown as functions of the cumulative enclosed mass $M(<r)$ of the rocky interior of embryos. The solidus pressures have been converted to mass by interpolation of the hydrostatic $P-m$ profiles. The dotted vertical line indicates the delimitation between the silicate mantle and the metal core, which is assumed to account for 32.5\% of the total mass following an Earth-like structure \citep{Sorokhtin_2011}.}

\label{fig:adiabat_vs_solidus}
\end{figure}

To address this concern, we compute interior thermal profiles for the rocky interior of early Earth embryos and quantify when the silicate melting curve is crossed. Using as outer boundary conditions the rocky interior mass $M_{\rm rock}$, the envelope-mantle surface pressure $P_0 \lesssim 0.1 {\rm GPa}$ found in Figure \ref{fig:example_envelope} and a set of surface temperatures, we integrate the structure equations (Equations \ref{eq:dM}-\ref{eq:dT}) radially inward throughout the mantle of embryos. The adiabatic cooling gradient of the mantle is set to

\begin{equation}
    \label{eq:nabla_ad_rock}
    \nabla_{\rm ad} = \frac{\alpha P}{c_P \rho},
\end{equation}
where $\alpha = 3.304  \times 10^{-5} {\rm \  K}^{-1} + ( 0.742 \times 10^{-8} {\rm K^{-2}})T$ is the mantle thermal expansivity \citep{Su_2018} and $c_P = 1.2561 {\ \rm kJ/kg/K}$ is the heat capacity of the mantle \citep{Elkins-tanton_2008}. For simplicity of this estimate, we assume complete differentiation of Fe into an iron core with the same core mass fraction $M_{\rm core}/M_{\rm rock} = 32.5\%$ as the Earth's core \citep{Sorokhtin_2011}.

The resulting interior profiles are compared in Figure \ref{fig:adiabat_vs_solidus} with the solidus of the silicate mantle. We use peridotite data for $P\leq 10 {\rm GPa}$ from \cite{Hirschmann_2000} to represent an Earth-like upper mantle and complementary data from \cite{Andrault_2011} for a higher-pressure chondritic deep mantle. We find that $0.3 M_\oplus$ embryos sustain surface magma oceans until they cool down to temperatures $T_0\lesssim 1600{\rm K}$. However, $\gtrsim 20 \%$ of the mantle is expected to solidify after an envelope-mantle boundary temperature of $\approx 1800{\rm K}$ is reached. Repeating the comparison for $0.2M_\oplus$ and $0.4M_\oplus$ embryos gives a similar lower bound on the minimal surface temperatures for which a liquid mantle can be assumed to efficiently deliver dissolved volatiles to the deep mantle.
Consequently, we build our dissolution model such that volatiles are delivered through the entire silicate mantle when $T_0 \geq T_{\rm melt}$, while bulk mantle reservoirs plateau once $T_0 < T_{\rm melt}$.\footnote{ By construction, the base temperature of the atmosphere $T(R_{\rm rock})$ is presumed to smoothly transitioned to the surface temperature of the melt $T_0$ over the entire accretion process. This assumption is valid until the solidification front of the magma reaches the surface \citep{Schaefer_2016}. In reality, the mantle thermally responds to changes in $T_0$, over timescales discussed in Section \ref{subsec:limations}. Even if envelope-mantle thermal equilibrium were to be disrupted, the resulting conductive boundary layer would be effectively non-existent \citep{Solomatov_2015}, with a thickness of $\delta \sim \frac{4 \pi R_{\rm rock}^2k_{\rm cond}}{L} ( T_0 - T(R_{\rm rock})) \sim 0.3 {\rm mm} \frac{( T_0 - T(R_{\rm rock}))}{1000{\rm K}}$ set by the thermal conductivity $k_{\rm cond}\approx 1.5 {\rm \ W/m/K}$ of the melt \citep{Hsieh_2024} and the luminosity $L \sim 10^{21}{\rm W}$ found in our calculations.}

We use a $^{22}$Ne elemental abundance in the primordial nebula of $n_{\rm ^{22}Ne}=(7.24 \pm 0.83) \times 10^{-6}$ obtained via three-dimensional (3D) radiative-spectroscopic hydrodynamical analyses of the solar photosphere \citep[see][their Table 2]{Asplund_2021}. 
In theory, the differential solubility of each atmospheric species, mainly that of H and He which dominate the primordial gas, along with rapid mixing timescales in the deep convective envelope could alter atmospheric composition.
We verify a posteriori that the mass of volatiles infused in the mantle throughout the entire accretion process is small compared to the total atmospheric budget  for H ($\lesssim 3\%$), He ($\lesssim 3\%$), and metals ($\lesssim 0.1\%$). Subsequent changes in the atmospheric composition can therefore be ignored with small fractional changes of ${\mathrm \Delta X/X\approx \Delta Y/Y \lesssim 0.06\%}$ and $\Delta Z/Z \lesssim 3\%$.
This verification validates the assumption of a constant envelope chemical composition and simplifies the dissolution calculation. 

Regarding the solubility constant, \cite{Iacono-Marziano_2010} characterized the linearity of the solubility of He, Ne and Ar with pressure in silicate melts, in agreement with the previous work of \cite{Jambon_1986}. From the basaltic lava samples of the 2002 Mt.~Etna eruption with $\mu^{M_{\rm Basalt}} = 64.13 \pm 0.38 \mathrm{\ g/mol}$ (see their Table 1), \cite{Iacono-Marziano_2010} developed a model to compute the solubility constant of Ne (along with He and Ar) within 22\% of uncertainty as a function of temperature and pressure, representative of the behavior of gas in basaltic melt up to 3 GPa and 1600 °C. We use their provided worksheet (see their supplementary data) to compute $k_{\rm Ne}$ as a function of $T_0$ and $P_0$.\footnote{For reference, the behavior of the solubility constant is well approximated by $k_{\rm Ne} \approx (9.1 \pm 2.0) \times 10^4 \mathrm{MPa} \Big(\frac{T_0}{1800 \mathrm{K}} \Big)^{-1.762}$ over the surface temperature and pressure ranges covered in our calculations. These values are in 1-$\sigma$ agreement with \cite{Lux_1987} and \cite{Miyazaki_2004} who respectively reported solubilities of $S_{\rm Ne} \equiv (\mu^{M_{\rm Basalt}}k_{\rm Ne})^{-1} \approx 13.4 \times 10^{-9} {\rm mol/(g \ atm)}$ at 1350$^{\circ} $C and $S_{\rm Ne} = 11-14 \times 10^{-9} {\rm mol/(g \ atm)}$ at 1300$^{\circ} $C.} In lack of complementary calculations for higher temperatures ($T_0 \gtrsim 1600\mathrm{^{\circ} C} = 1873.15\mathrm{K}$), we extrapolate the solubility model of \cite{Iacono-Marziano_2010} using their Equations (8-10) to the higher values of $T_0\lesssim 2500 \mathrm{K}$ retrieved at the bottom of our computed primordial envelopes (see Figure \ref{fig:example_envelope}). The impact of this extrapolation is minimal as we find that most of dissolved volatile content ($\gtrsim 80\%$) is delivered as the mantle nears its solidification ($1800 {\rm K} \lesssim T_0 \lesssim 1900 {\rm K}$).

After formation over $\sim4.5$ Gyr of evolution of the Earth, the dissolved gas can be outgassed by the continuous motion of tectonic plates, mantle convection, and volcanic activity, which are expected to drive a steady outgassing rate of volatiles from the mantle \citep{Lupton_1975,Farley_1995,Bender_2008,Bianchi_2010,Moreira_2013,Vayrac_2025}.
To account for these effects, we adopt the best-fit forward model of \cite{Parai_2022} to calculate the fraction of the noble gas content of the mantle that was lost due to outgassing. They argued that even though the lower mantle has been more efficient at retaining its initial gas than the upper mantle, it still suffered significant degassing. From their best-fit parameters (see their Figure S6) for the plume mantle, we find that the initial deep mantle Ne reservoir was $\approx 18.7$ times more enriched relative to the present-day budget.
We consequently multiply the quoted concentration of the deep mantle $^{22}$Ne reservoir \citep{Marty_2012,Halliday_2013} by $\approx 18.7$ to correct for mantle outgassing in Section \ref{sec:results}. In Section \ref{subsubsec:cohesive_formation_pathway}, we consider additional loss of deep mantle volatiles during the giant impact era, but conclude that any ejected material would most likely by reaccreted and redissolved in the interior.
   
\section{Results} \label{sec:results}

With the one-dimensional differential system presented in Section \ref{subsec:gas_accretion}, we simulate the envelope growth of planetary embryos ranging from $M_{\rm rock} = 0.1 M_\oplus$ to $M_{\rm rock} = 0.5 M_\oplus$. 
We do not consider embryos of higher masses because of the evidence of the Earth's formation via giant impact(s) post-nebular dispersion \citep{Tucker_2014}, which requires proto-planets of mass $M_{\rm rock} \leq 0.5 M_\oplus$. 
We explore depletion factors of $\log f_{\rm dep} \in [0,-1,-2,-3,-4,-5,-6]$ (see Equation \ref{eq:rho_MMSN}) to explore formation at later stages of the lifetime of the disk, as the latter gradually dissipates. 

We find that dissolution of nebular gas is only possible in the early stages of envelope formation while the mantle stays molten (i.e. the red dashed line for the example case of Figure \ref{fig:example_envelope}), as the surface temperature of the envelope-mantle boundary quickly reaches the solidification temperature of the mantle (i.e. $T_0 \approx T_{\rm melt} $).
We therefore stress the importance of adopting a careful treatment of the cooling of primordial envelopes to avoid overestimating the dissolved concentration of noble gases in the mantle.
For example, if one were to calculate the dissolved fraction of $^{22}$Ne in the mantle (using Equation \ref{eq:Henrys_law}) based on the near-isothermal end state of the envelope shown in the top panel of Figure \ref{fig:example_envelope}, the primordial Ne budget of the mantle would be overestimated by several orders of magnitude due to the large increase of $P_0$ from $\approx 10^{-5}$ MPa to $\approx 10$ MPa during the gas accretion stage.

Our gas accretion calculation also argues that the accretion timescale of volatiles $t_{\rm acc}$ over which $^{22}$Ne can be dissolved in the deep mantle is very short (e.g. $\lesssim 10^{-4}  \mathrm{\ yr}$ for the $f_{\rm dep} =10^{-2}$, $M_{\rm rock} = 0.3 M_\oplus$ case of Figure \ref{fig:example_envelope}).
We therefore cannot use this model to establish clear constraints on the formation timescales of primordial envelopes in the disk or on the lifetime of the nebula itself.
For example, embryos accreting gas from a disk with $f_{\rm dep} = 10^{-2}$ on any timescale $\Delta t\gtrsim 10^{-3} \  \mathrm{yr}$ on top of a $0.3 M_\oplus$ core would contain similar dissolved concentrations of $^{22}$Ne in their mantle. 
The short values of $t_{\rm acc}$ before which the base of primordial envelopes reaches the melting temperature of basalt raise questions as to whether the assumption of rapid mixing and thermal balance between the molten mantle and the forming gas envelope is reasonable. In light of those concerns, we present the results of the adopted model in Section \ref{subsec:Ne_results} and assess in Section \ref{subsec:limations} the physical consistency of the rocky interior of Earth embryos cooling and mixing over the timescales required by our solutions.

\subsection{Accretion of primordial envelopes}
\label{subsec:Ne_results}

\begin{figure*}[t]
\subfigure[]{\includegraphics[width=9cm]{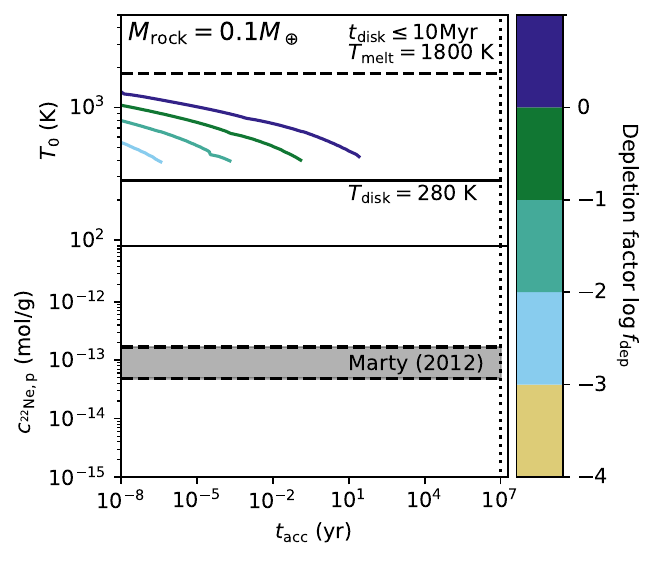}\label{subfig:Ne_0.1}}
\subfigure[]{\includegraphics[width=9cm]{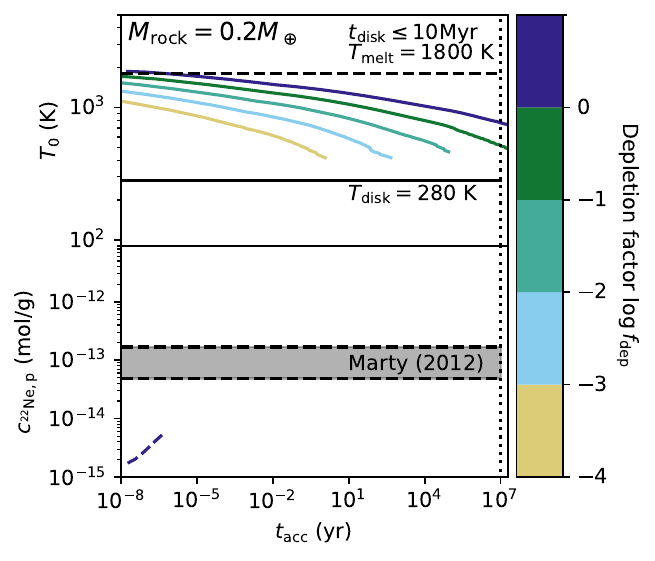} \label{subfig:Ne_0.2}}
\\
\subfigure[]{\includegraphics[width=9cm]{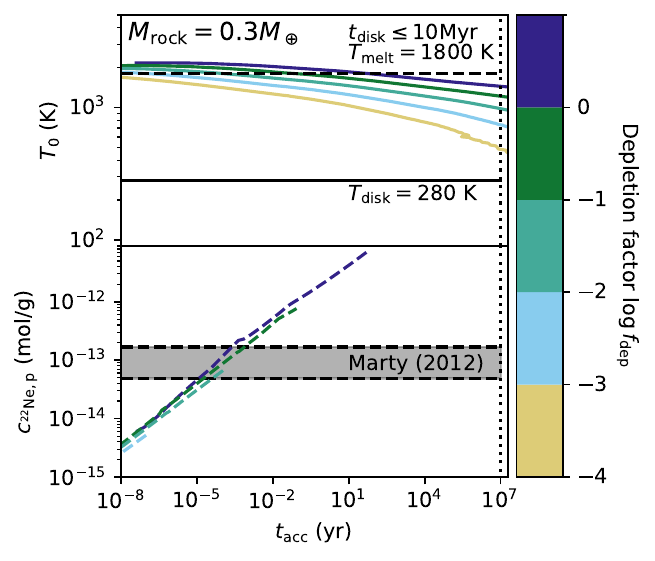}\label{subfig:Ne_0.3}
}
\subfigure[]{\includegraphics[width=9cm]{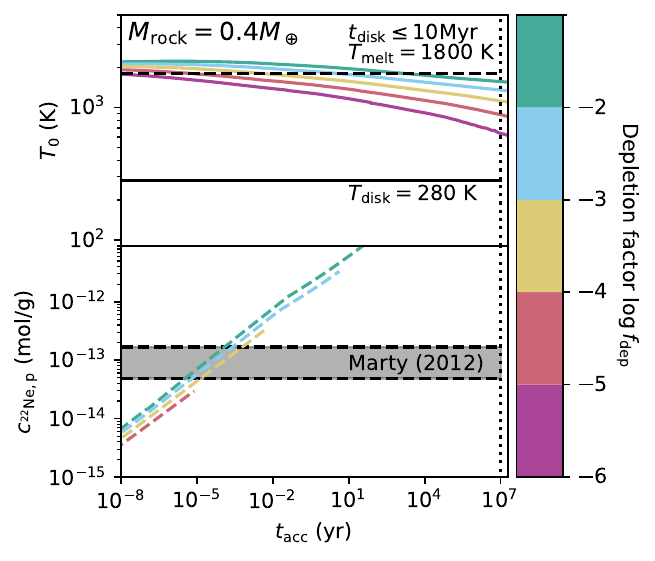}\label{subfig:Ne_0.4}}

\caption{Dissolution calculation of the concentration of primordial $^{22}$Ne captured at the molten surface of magma oceans on Earth embryos embedded in the solar nebula. Subfigures (a), (b), (c) and (d) present our results for protocore masses of $0.1,0.2,0.3,0.4M_\oplus$, respectively. For each case, the upper panel shows the temperature $T_0$ of the envelope-mantle boundary as a function of gas accretion time $t_{\rm acc}$ and the lower panel the resulting concentration of primordial $^{22}$Ne dissolved in the silicate interior $c_{\rm ^{22}Ne,p}$ before the mantle solidifies at a melting temperature $T_{\rm melt}\approx 1800{\rm K}$. The shaded areas represent the target concentration required to explain the present-day budget of the deep mantle, as constrained by \cite{Marty_2012} and corrected by a factor of $\approx 18.7$ to account for mantle outgassing \citep{Parai_2022}. The solid black line indicates the temperature $T_{\rm disk}$ of the minimum-mass solar nebula (MMSN) of \cite{Hayashi_1981} and the black dashed line the melting temperature $T_{\rm melt}$ of the basaltic mantle (Section \ref{subsec:limations}), required at the surface for the dissolution of primordial gas in the interior. The upper bound on the lifetime of the disk \citep{Mamajek_2009} is indicated with a dotted vertical line, corresponding to the maximal possible time after which primordial accretion must come to an end. Different colors account for different depletion factors $f_{\rm dep} \leq 1$ of the gas density of the MMSN disk.}

\label{fig:Ne_dissolved}
\end{figure*}

Following the procedure of Section \ref{subsec:dissolution}, we calculate the concentration $c_{\rm ^{22}Ne,p}$ of primordial $^{22}$Ne dissolved in the mantle at surface pressure $P_0$ in the presence of a molten mantle ($T \geq 1800 \mathrm{K}$) during early accretion. Figure \ref{fig:Ne_dissolved} presents our findings for embryo mass of $M_{\rm rock} = [0.1,0.2,0.3,0.4] M_\oplus$.

We find that protocores of mass $M_{\rm rock} \lesssim 0.1 M_\oplus$ are so light that they only accrete limited amount of gas, which does not generate enough accretional heat to sustain magma oceans at their surface (see Figure \ref{subfig:Ne_0.1}).
For larger embryos with surface temperatures that allow envelope-mantle mixing, we aim to compare our predicted Ne reservoirs with the present-day budget of the deep mantle, correcting for mantle outgassing. Using a global mass balance analysis of the Earth's inventory of $^{40}$Ar, \cite{Marty_2012} derived the volatile composition of the bulk silicate mantle (i.e. the non-degassed mantle) and constrained its concentration of $^{22}$Ne to $c_{\rm ^{22}Ne,DM} =(5.8 \pm 3.2) \times 10^{-15} \ \mathrm{mol/g}$. We adopt this concentration for deep mantle reservoirs, but also note that similar calculations by \cite{Halliday_2013} predict a lower Ne budget by $\approx 1 \sigma$, with the discrepancy arising from the uncertainty in extrapolating the compositions of OIB samples to the deep mantle. 
Guided by the results of \cite{Williams_2019}, we assume for simplicity that this entire Ne reservoir in the deep mantle is of nebular origin. This corresponds, after accounting for mantle degassing, to an initial concentration of $c_{\rm ^{22}Ne,p}= (1.08 \pm 0.60) \times 10^{-13} \ \mathrm{mol/g}$ dissolved during the accretion of a primordial envelope, which we set to our target value. Secondary contributions from accreted chondrites and/or subducted atmospheric gas are expected to be minor ($\lesssim 10-15\%)$ \citep{Peron_2022}, and are discussed in Section \ref{sec:Noble_gas_budget}. 
 
Comparing with the reference from \cite{Marty_2012} in Figure \ref{fig:Ne_dissolved}, we first find that the nebular Ne reservoir cannot be explained by accretion onto $\lesssim 0.2 M_\oplus$ Earth embryos. Their mantle solidifies  too quickly to reach an agreement with the measurements of \cite{Marty_2012}, with a final value of $c_{\rm ^{22}Ne,p}$ lower than the target by an order of magnitude. Instead, Figure \ref{subfig:Ne_0.3} shows that protocores of mass $M_{\rm rock} = 0.3 M_\oplus$ emerging in a depleted nebula with $10^{-2} \lesssim  f_{\rm dep} \lesssim 10^{-1}$ would explain the present bulk mantle budget. Such solutions for low values of $f_{\rm dep}$ in protoplanetary disk correspond to a late stage formation scenario of rocky protocores in the primordial gas \citep{Lee_2016}. The dissolution timescale required to reach the $c_{\rm ^{22}Ne,p}= (1.08 \pm 0.60) \times 10^{-13} \ \mathrm{mol/g}$ target is very short ($\lesssim 0.01$ yr) and justifies the use of a time-independent disk profile (Equation \ref{eq:rho_MMSN}), since the disk gas density would not evolve substantially over this short time frame. 

Finally, for higher embryo masses of $\geq 0.4 M_\oplus$, Figure \ref{subfig:Ne_0.4} displays excesses in $c_{\rm ^{22}Ne,p}$ by at least an order of magnitude unless embryos emerged in a $f_{\rm dep} \lesssim 10^{-4}$ gaseous nebula.
We discuss in Section \ref{subsec:Earth_final_assembly} how a surplus of dissolved volatile content in the deep mantle post-accretion is unlikely to be ejected via subsequent mass loss events such as giant impacts. 
Additionally, we expect minimal delivery of $^{22}$Ne from the molten silicate interior of embryos to segregated metal cores due to the low Ne metal/silicate coefficients of $D_{\rm Ne} \lesssim 0.1\%$ at core-mantle boundaries \citep{Wang_2022}.\footnote{ It has been suggested that primitive He and Ne is leaking from the Earth's core \citep[e.g., ][]{Vogt_2021,Olson_2022}, due to the decrease in their partition coefficients with core-mantle boundary temperature \citep[see Figure 2 of][]{Wang_2022}. However, the low value of $D_{\rm Ne}\lesssim 0.1\%$ during formation places an upper bound of $c_{\rm ^{22}Ne,p}/1000 \sim 10^{-16} {\rm mol/g}$ on the initial Ne core reservoir, that is a minimal fraction of the current bulk mantle concentration of $c_{\rm ^{22}Ne,DM} =(5.8 \pm 3.2) \times 10^{-15} \ \mathrm{mol/g}$. Therefore, whether or not an outflow of nebular Ne from the core exists, its effect on deep mantle concentrations would be minimal.}
The values of $c_{\rm ^{22}Ne,p}$ dissolved in Figure \ref{fig:Ne_dissolved} are therefore directly related to the bulk mantle concentrations prior to the steady degassing of the mantle over $\sim$Gyr timescales. 
Consequently, the accretion of a primordial envelope from a depleted $10^{-4} \lesssim  f_{\rm dep} \lesssim 10^{-1}$ gas disk atop $\gtrsim 0.4M_\oplus$ protocores in Figure \ref{subfig:Ne_0.4} overshoots the expected $^{22}$Ne concentration from the current Ne budget of the deep mantle.

The possibility of a $0.4 M_\oplus$ solution in a $f_{\rm dep} \lesssim 10^{-4}$ gas disk must also be addressed. In theory, the accretion of gas envelope in low-density protoplanetary disks is possible for gas densities $\rho_{\rm disk} \gtrsim 10^{-8}\rho_{\rm MMSN}$ \citep{Lee_2018}, and lower values of $f_{\rm dep} < 10^{-4}$ are explored in Figure \ref{subfig:Ne_0.4}.
For $f_{\rm dep} < 10^{-5}$, we do not find a solution that matches the target Ne concentration.
In $f_{\rm dep} < 10^{-6}$ gas disks, the bottom of the envelope does not reach the melting temperature of the mantle $T_{\rm melt} \approx 1800{\rm K}$.
There may exist a marginal solution in the range of depletion factor $10^{-5} \lesssim f_{\rm dep} \lesssim 10^{-4}$, but not below.
The challenge of reconciling an embryo mass of $M_{\rm rock} = 0.4 M_\oplus$ with the primordial $^{22}$Ne reservoir establishes an upper limit on the mass of Earth embryos that must have formed in the solar nebula.

In summary, our model argues that a nebular origin for Ne in the deep mantle requires the emergence of protocores of masses $M_{\rm rock} \sim 0.3-0.4 M_\oplus$ in a solar nebula of gas density depleted by $f_{\rm dep}\sim 10^{-4}-10^{-1}$. The Earth embryos would accrete a thin ($M_{\rm env}/M_{\rm rock} \lesssim 10^{-5}$) primary gas envelope, rapidly dissolving the measured primordial $^{22}$Ne concentration in the deep mantle.
Our results exclude $ M_{\rm rock} \leq 0.2 M_\oplus$ solutions for which envelope-mantle surface temperatures $T_0$ are too low to facilitate the efficient delivery of $^{22}$Ne to the mantle of embryos. The $ M_{\rm rock} > 0.4 M_\oplus$ case is also excluded since either the surface is not warm enough ($T_0<T_{\rm melt}$) to sustain a molten interior or high surface pressures $P_0$ lead to an overestimation of the deep mantle primitive $^{22}$Ne reservoirs.  Consequently, the constraint on the mass of early Earth embryos embedded in the primordial nebula is tight and their emergence most likely occurred in the later stages of the gas disk.
The lower bound of $M_{\rm rock} > 0.2 M_\oplus$  is in agreement with the analogous study of the imprint of a nebular atmosphere on the interior of Mars-mass embryos of \cite{Jaupart_2017}.

\subsection{ Cooling and mixing of the mantle}

\label{subsec:limations}

The gas accretion model presented in Section \ref{sec:methods} is dependent on the key assumption that volatiles dissolved at the surface of magma oceans are instantly delivered to the deep mantle when the latter is liquid (i.e. when $T_0 > T_{\rm melt} \approx 1800{\rm K}$ as shown in Figure \ref{fig:adiabat_vs_solidus}). Two criteria must be met by the mantle-envelope system for this simplification to be physically consistent.

First, even in an entirely molten silicate mantle, instantaneous chemical mixing is only justifiable if the radial circulation of the molten mantle is faster than the rate at which volatiles are delivered to the interior. 
To assess this first criterion, we consider the mantle convective velocity $v_{\rm con}$, as given by \citet{Priestley_1957,Priestley_1959,Solomatov_2000b}:

\begin{equation}
    \label{eq:v_mix_formula}
    v_{\rm con} = 0.6 \Big( \frac{\alpha g F_0 \Delta R }{{\bar \rho} c_{P}} \Big)^{1/3}.
\end{equation}
Here, $g = 9.82 {\ \rm m/s^2} \Big( \frac{M}{M_\oplus} \Big)^{1/3}$
is the standard surface acceleration of embryos of constant bulk density $\bar{\rho} \equiv \frac{3M_\oplus}{4\pi R_\oplus^3} \approx 5500 {\ \rm  kg/m^3}$,\footnote{ While a bulk density of $\approx 5500{\rm \ kg/m^3}$ is high for Mars-sized embryos, we use this value for self-consistency with the rocky interior radius $R_{\rm rock} = R_\oplus \Big( \frac{M_{\rm rock}}{M_\oplus} \Big)^{1/3}$ adopted in Section \ref{subsec:gas_accretion}. This choice has little impact ($\lesssim 7\%$) on the estimate of Equations (\ref{eq:v_mix_scaling}) and (\ref{eq:tau_mix}).} $F_0 \equiv \frac{L}{4\pi R_{\rm rock}^2}$ is the cooling flux of the mantle driving convective flows and  $\Delta R \equiv R_{\rm rock} - R_{\rm core}$ is the radial extent of the mantle, which is the relevant convective length scale.
We equate the cooling of the young molten mantle to the atmospheric luminosity, implicitly assuming thermal equilibrium at the surface of the magma oceans. This assumption justifies the use of $L$ in Equation (\ref{eq:v_mix_formula}).
The core-mantle structure of Earth embryos is hardly constrained, and we thus simply estimate $\Delta R$ by scaling down the radial extent of the Earth reported by \cite{Young_1987},
assuming a constant bulk density $\bar{\rho}$, similar relative metal-silicate compositions and a complete differentiation of Fe to form an iron core:

\begin{equation}
    \label{eq:Delta_R__mantle}
    \Delta R \sim 0.45 R_\oplus \Big( \frac{M_{\rm rock}}{M_\oplus} \Big)^{1/3} = 0.45 R_{\rm rock}.
\end{equation}

For the embryos considered in this work, we find that convective velocities as given by Equation (\ref{eq:v_mix_formula}) are expected to be on the order of

\begin{equation}
    \label{eq:v_mix_scaling}
    v_{\rm con} \sim 7.3 {\rm \ m/s} \Big( \frac{L}{10^{21} {\ \rm W}} \Big)^{1/3}.
\end{equation}
which corresponds to a mixing timescale $\tau_{\rm mix} \equiv \Delta R/v_{\rm conv}$ of 

\begin{equation}
\label{eq:tau_mix}
    \tau_{\rm mix}  \sim 0.013 {\rm \ yr} \Big( \frac{10^{21} {\rm \ W}}{L} \Big)^{1/3} \Big( \frac{M_{\rm rock}}{M_\oplus} \Big)^{1/3}, 
\end{equation}
or $\sim 0.0084 {\rm yr}$ for the favored $\sim 0.3 M_\oplus$ solution with luminosity of the gas envelope $L \sim 10^{21} {\ \rm W}$ retrieved in our calculations in the end stage of dissolution (i.e. for $T_0\approx T_{\rm melt} = 1800 {\rm K}$).
Comparing $\tau_{\rm mix}$ with the short accretion timescales of $t_{\rm acc}\lesssim 0.002 {\rm \ yr}$ required by the solutions of Figure \ref{fig:Ne_dissolved} yields $t_{\rm acc}/\tau_{\rm mix} \lesssim 1/4$. The validity of the rapid mixing assumption is therefore questionable, necessitating follow-up geodynamical considerations, which is beyond the scope of this paper. We therefore proceed with our assumption with this caveat in mind.\footnote{We note that a departure from a purely dust-free opacity regime would validate the assumption of rapid mixing ($\tau_{\rm mix} \lesssim t_{\rm acc}$). The accretion time scales linearly with opacity ($t_{\rm acc} \propto L^{-1} \propto \kappa$) and the dependency of the mixing timescale estimated in Equation (\ref{eq:tau_mix}) on opacity is sub-linear ($\tau_{\rm mix} \propto L^{-1/3} \propto \kappa^{1/3}$). Therefore, an opacity enhancement from dust by a factor of 8 would be sufficient, which is well within reason considering that an ISM-like distribution of dust grains \citep{Ferguson_2005} increases the dust-free opacities of the envelopes of Figure \ref{fig:example_envelope} from $\sim 0.001 {\rm \ cm^2/g}$ to $\sim 1 {\rm \ cm^2/g}$. Implications of dusty envelope is discussed more in Section \ref{subsubsec:gas_rich_formation}.}

Second, the convective mantle velocity of Equation (\ref{eq:v_mix_scaling}) is sustained by the convective cooling of the mantle over timescales of 
\begin{equation}
    \tau_{\rm th} \sim \frac{M_{\rm mantle} c_p \Delta T}{L},
\end{equation}
where $M_{\rm mantle}$ is the total mass of the silicate mantle and $\Delta T$ is the difference between $T_0$ and the temperature of the deep mantle layers.  Assuming once more for simplicity that the interior of embryos follows an Earth-like structure with $M_{\rm mantle} \approx 67.5\% M_{\rm rock}$ \citep{Sorokhtin_2011}, the convective mixing of the mantle of the favored $\sim 0.3 M_\oplus$ embryos with retrieved luminosity values of $L\sim 10^{21} {\rm \  W}$ is sustained over

\begin{equation}
\label{eq:tau_th_scaling}
    \tau_{\rm th} \sim 46 {\rm \ yr} \frac{\Delta T}{1000 {\rm K}}.
\end{equation}
Over this timescale, the temperature gradient is erased and convection halted. It is therefore unrealistic to assume that $^{22}$Ne could be mixed to the deep mantle after a time $\tau_{\rm th}$. 

From Figure \ref{fig:adiabat_vs_solidus}, we can approximate a value of $\Delta T \gtrsim 1000 {\rm \ K}$ and obtain a lower bound of $\tau_{\rm th} \gtrsim 46 {\rm \ yr}$, longer than $t_{\rm acc} \lesssim 10^{-2} {\rm \ yr}$. The interior temperature gradient thus persists over the entire dissolution stage and sustains the mantle convective velocities of Equation (\ref{eq:v_mix_scaling}). We verify a posteriori that the interior thermal structure of Figure \ref{fig:adiabat_vs_solidus} would favor turbulent flow, as can be assessed from the Rayleigh number of the melt

\begin{equation}
    \label{eq:Ra_def}
    {\rm Ra} \equiv \frac{\alpha \rho^2 g c_P (R_{\rm rock} - r)^3}{\eta k_{\rm cond}} \Delta T,
\end{equation}
where $k_{\rm cond}$ and $\eta$ are the thermal conductivity and dynamical viscosity of the liquid silicate. Above a critical value of ${\rm Ra_c} \sim 1000-2000$, magma oceans are unstable to convection \citep{Koschmieder_1993}. Using typical values of $k_{\rm cond}\approx 1.5 {\rm \ W/m/K}$ \citep{Hsieh_2024} and $\eta \sim 0.1 {\rm \ Pa \cdot s}$ \citep{Solomatov_2015}, Equation (\ref{eq:Ra_def}) gives ${\rm Ra} \gtrsim 10^{20} \gg {\rm R_c}$, characteristic of a strong convective regime.
As such, we expect volatiles to be efficiently transported through the mantle over the accretion and dissolution timescale of $t_{\rm acc} \lesssim 0.002 {\rm \  yr}$ retrieved in our model (Figure \ref{fig:Ne_dissolved}).

\section{ Earth formation history}
\label{sec:formation}

\subsection{Embryo formation in the primordial nebula}
\label{subsec:embryo_formation}

We have so far worked under the assumption that fully assembled Earth embryos accreted their primordial gas envelope in-situ at $a=1$ au in the later stages of the solar nebula, at a time when dust grains no longer contributed to the opacity.
This choice allows us to present a self-consistent solution for which $\sim 0.3-0.4 M_
\oplus$ embryos build their envelope in the gas-poor ($f_{\rm dep} \sim 10^{-4}-10^{-1}$) environment of the nebula near its dispersal. Here, we assess the validity of this primary assumption and investigate how departures from it may impact our results. 

\subsubsection{Consistent formation scenario in a late-stage nebula}
\label{subsubsec:late_stage_formation}

Solid growth of Earth embryos can proceed through planetesimal accretion \citep{Safronov_1969,Safronov_1972,Hayashi_1985,Wetherill_1989,Ida_1993}, where km-scale bodies collide and merge, and/or through pebble accretion \citep{Ormel_2010,Johansen_2010}, where mm-cm sized particles drifting inward via aerodynamic drag in the disk are efficiently captured by individual accretors.
Since there remains debate as to which of these accretion mechanisms predominantly drives embryo formation in the early solar system \citep{Izidoro_2021,Onyett_2023,Morbidelli_2025}, we remain agnostic to the specific pathway by which Earth embryos were assembled. 

That said, it must be noted that both scenarios, along with the possibility of mergers via early giant impacts, can easily lead to the formation of $\sim 0.3-0.4M_\oplus$ Earth embryos at $\sim$1 au in the solar nebula within 1-5 Myr \citep{Weidenschilling_1997,Kokubo_2000,Chambers_2006,Walsh_2016,Walsh_2019,Izidoro_2021,Johansen_2021}.
Furthermore, migration in low density protoplanetary disks is expected to be inefficient for $\sim 0.1-0.5 M_\oplus$ proto-planets due to the limited torque from the gas \citep{Tanaka_2002}. Most formation models of the inner solar system suggest that the terrestrial planets formed largely in-situ, while the giant planets experienced the bulk of migration \citep{Hansen_2009,Jacobson_2014, Kaib_2016,Brasser_2025,Morbidelli_2025}.
It is therefore fairly reasonable to consider the formation of $\sim 0.3-0.4 M_
\oplus$ ancestors of the Earth near $\sim$1 au within the first few Myr of the solar system.

The remaining key question to be answered is the state of the solar nebula at the time those embryos emerged, which is challenging to determine.
Observationally, the lifetime of protoplanetary disks around Sun-like stars has been inferred to be 1-10 Myr \citep{Mamajek_2009,Alexander_2014,Michel_2021} from observations of young stellar objects. 
For the Sun, modern formation models of the gas giants \citep[e.g.,][]{Benvenuto_2009,DAngelo_2021} require the presence of the gaseous nebula up to $\sim 3$ Myr into the formation of the solar system.

Although remnants of the primordial nebula in the Earth's deep mantle do not directly constrain the disk dissipation timescale, our results argue that the emergence of $\sim 0.3-0.4 M_\oplus$ Earth embryos must have coincided with the dispersal of the gas disk. Indeed, primordial gas must be present, but not too much $(10^{-4} \lesssim f_{\rm dep} \lesssim 10^{-1}$) to explain the concentration of primordial gas in the rocky interior of embryos.

The formation of primordial atmospheres in a gas-poor nebula is consistent with our use of dust-free opacity grids.
As the disk evolves on $\sim$Myr timescales, solid dust grains in the disk either grow into bigger objects (e.g., pebbles, planetesimals, embryos, \citealt{Ormel_2014,Mordasini_2014}), drift radially inward toward the Sun \citep{Laibe_2012}, or are blown out via photoevaporation and disk winds \citep{Rodenkirch_2022}. 
Furthermore, dust grains within primordial atmospheres can settle (``rain out") toward deeper layers, thereby favoring dust-free regimes \citep{Ormel_2014,Mordasini_2014b}.
In the absence of the opacity contribution of dust grains, thin primordial envelopes are rapidly accreted as they efficiently cool, quickly delivering volatiles to the rocky interior before mantle solidification. 

\subsubsection{Marginal solution in a gas-rich nebula}
\label{subsubsec:gas_rich_formation}

It must be noted that a marginal alternative scenario of embryos of intermediary mass (e.g. $0.25 M_\oplus$) between the cases of the upper panel of Figure (\ref{fig:Ne_dissolved}) may provide a solution in a gas-rich ($f_{\rm dep}=1$) disk and merits consideration. Such a solution would be at odds with our dust-free opacity assumption, as we may expect disk opacities to gradually transition from ``dusty" to dust-free regimes as the nebula dissipates. To evaluate the validity of this result, we repeated the gas accretion calculation of Section \ref{subsec:gas_accretion} with the dusty opacity grids of \cite{Ferguson_2005}, which assume an interstellar medium size distribution of dust grains.\footnote{The \citet{Ferguson_2005} opacity table cuts off below $T  = 500{\rm K}$, where we extrapolate using a $\kappa \propto T^2$ scaling.}  

We find that the thermal $(P,T)$ structure of individual snapshots is largely unaffected by the change in opacity grid, resulting in no significant variations in the dissolved $^{22}$Ne concentrations reported in Figure \ref{fig:Ne_dissolved}. 
As the radiative outer layer extends inward, the dusty and dust-free envelopes transition to the deep layers following the same inner adiabat, resulting in similar $P_0$ values.

Unfortunately, the behavior of $\kappa(r)$ in dusty envelopes is non-monotonic, arising from significant opacity drops due to the sublimation of different grain species for $1500{\rm K} \lesssim T \lesssim 2500{\rm K}$.
This complex opacity profile leads to deviations from the simple envelope structure (innermost convective + outer radiative layers) of dust-free envelopes presented in Figure \ref{fig:example_envelope}, with additional inner and intermediary radiative shells. As such, the major assumption in our gas accretion model of a spatially constant luminosity fails for dusty envelopes, with large neglected contributions from inner radiative layers in Equation (\ref{eq:L_negl}) (i.e. $L_{\rm negl} \gtrsim L$). We therefore cannot use our scheme to resolve the precise time evolution of primordial dusty envelopes, as is done for dust-free regimes in Figure \ref{fig:example_envelope}.

The timescale required to dissolve the required $^{22}$Ne in a dusty environment can nonetheless be estimated by comparing the thermal structures between dusty and dusty-free envelopes.
As illustrated in Figure \ref{fig:example_envelope}, the thin primordial envelopes accreted by small embryos lie in a temperature regime ($T\lesssim 2500{\rm K}$) where H$^-$ ions do not dominate the opacity. When present, dust grains therefore control the opacity, with dusty values being roughly three orders of magnitude higher than molecular opacities. 
In more opaque environments, planetary envelopes cool at a linearly slower rate since $L\propto \frac{1}{\kappa}$ (see Equation \ref{eq:grad_rad}). 
The formation of the primordial envelope and the subsequent transport of volatiles to the mantle thus progress more slowly, stretching the timescale of our solution (Figure \ref{subfig:Ne_0.3}) from only 3 hours for a dust-free envelope to $\sim 125$ days for a dusty equivalent.

Overall, the presence of dust grains in the early nebular gas would not alter the concentrations displayed in Figure \ref{fig:Ne_dissolved}, but only slow the accretion and dissolution processes.
Consequently, embryo masses of $M_{\rm rock} \leq 0.2 M_\oplus$ are also ruled out for dusty gas accretion and the solutions displayed in Figures \ref{subfig:Ne_0.3} and \ref{subfig:Ne_0.4} remain valid in the presence of dust grains.
We note that a solution in an early gas-rich disk ($f_{\rm dep }=1$) with dusty opacities may exist for an embryo mass of $0.2 M_\oplus < M_{\rm rock} < 0.3 M_\oplus$.

\subsection{Final assembly of the Earth}
\label{subsec:Earth_final_assembly}

\begin{figure*}[t]
    \centering
    \includegraphics[width=\linewidth]{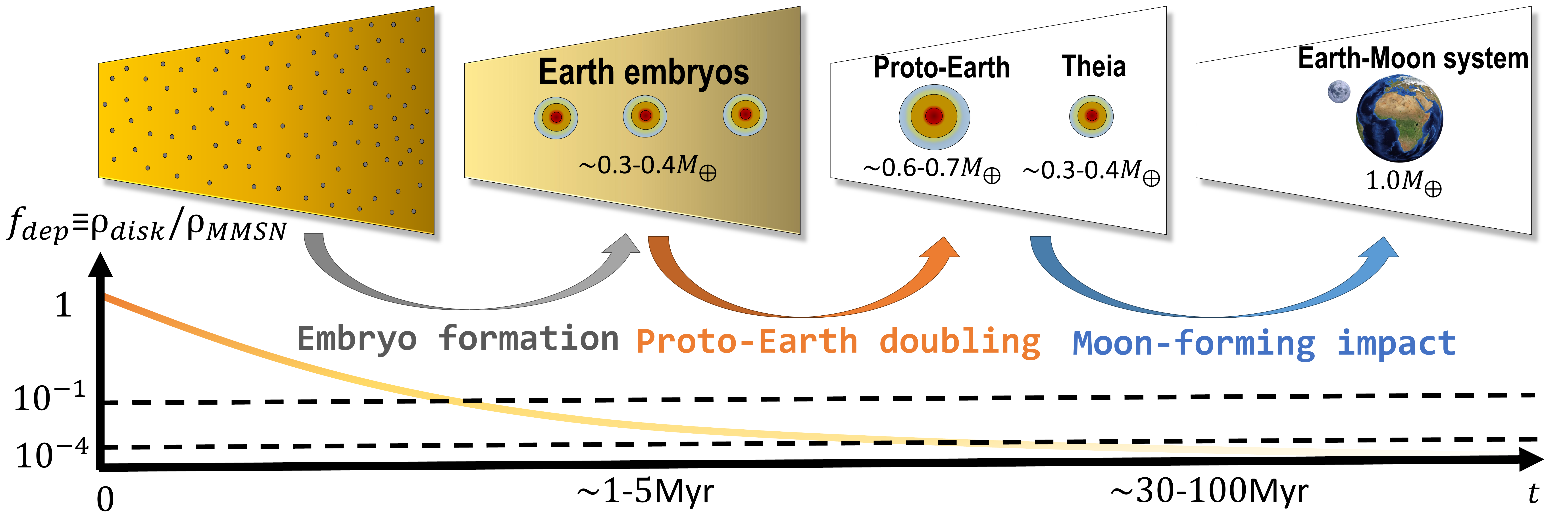}
    \caption{Schematic illustration of the favored Earth formation scenario implied by our results. Top: From left to right, we show the different formation stages of the Earth from a side-view of a truncated solar nebula, which is initially rich in gas (yellow) and dust (gray dots). As the disk progressively dissipates, the coagulation of solids (see Section \ref{subsubsec:late_stage_formation}) leads to the formation of a set of three $\sim 0.3-0.4 M_\oplus$ embryos, displayed with cyan, orange and red circles representing the layered embryo structure of Figure \ref{fig:structure_diagram}. Following the dispersal of the disk, embryo mergers are enabled, resulting in the formation of the proto-Earth via the doubling (orange arrow) of two $\sim 0.3-0.4 M_\oplus$ embryos. The proto-Earth later collides (blue arrow) with the remaining embryo (i.e. Theia) to form the final Earth-Moon system. Bottom: Time evolution (yellow curve) of the volume gas density $\rho_{\rm disk}$ of the solar nebula beginning from a full minimum-mass solar nebula density $\rho_{\rm MMSN}$, as parametrized by the parameter $f_{\rm dep}$ of Equation (\ref{eq:rho_MMSN}). Key values of $f_{\rm dep} = 10^{-1}$ and $f_{\rm dep} = 10^{-4}$ setting the boundaries between the gas-rich, gas-poor and dispersed stages of the disk are shown with dash horizontal lines.}
    \label{fig:favored_formation_scenario_impacts}
\end{figure*}

Thus far we have focused on the assembly of $\leq 0.4 M_\oplus$ planetary embryos in the solar nebula.
It is widely understood that the final coalescence of the Earth occurred via giant impacts \citep{Meier_2014}, where like-sized proto-planets collide and merge \citep[e.g., ][]{Agnor_1999}. 
This section links the formation of Earth embryos to the planet's present state and provides a unified picture of its evolution in Figure \ref{fig:favored_formation_scenario_impacts}.

\subsubsection{Evidence of giant impacts}
\label{subsubsec:impact_constraints}

According to the Giant Impact Hypothesis \citep{Hartmann_1975,Cameron_1976}, the formation of the final Earth-Moon system stems from a catastrophic collision of a young ``proto-Earth" with a lower-mass protoplanet often referred to as ``Theia" \citep{Halliday_2000,Quarles_2015,Kaib_2015}.
Evidence supporting this theory includes the spin and orbital properties of the Earth-Moon system being consistent with a massive collision \citep{Canup_2008}, the volatile depletion of the Moon implying high-temperature processing of the ejected material \citep{Day_2014}, and the isotopic similarities between the Earth and the Moon \citep{Wiechert_2001}.
Converging lines of reasoning date this final assembly to $\sim30-100$ Myr after the formation of Calcium-Aluminum-rich Inclusions (CAIs),\footnote{In this work, we do not distinguish between the onset of the solar system formation and the formation of CAIs, which are its earliest solids, and adopt this time as $t=0$.} using isotopic measurements in meteorites as chronometers \citep{Kleine_2009,Rudge_2010}, isotopic analyses of the Earth and the Moon \citep{Halliday_2008}, and dynamical models of their formation \citep{Agnor_1999,Raymond_2009}.

Details of the proto-Earth-Theia collision have been investigated by many studies \citep[e.g.][]{Canup_2004,cuk_2012,Canup_2012}.
The original idea of a ``Moon-forming" impact describes a collision between a proto-Earth of mass $M_{\rm PE}\approx 0.9 M_\oplus$ and a Mars-sized Theia of mass $M_{\rm Theia}\approx 0.1 M_\oplus$ \citep{Hartmann_1975,Cameron_1976,Canup_2004}.
In the classical scenario with an impactor-to-target mass ratio of $\phi \equiv  M_{\rm Theia}/M_{\rm PE} \approx 1/9$, the Moon forms from an orbiting silicate disk primarily originating from the impactor's mantle, while the Earth's composition remains consistent with the original proto-Earth \citep{Canup_2001}.

This classical picture expects distinct compositions between the Earth and the Moon, but is challenged by their isotopic similarities in oxygen \citep{Wiechert_2001}, chromium \citep{Lugmair_1998}, titanium \citep{Zhang_2012} and tungsten \citep{Dauphas_2017,Kruijer_2017}.
Invoking the coalescence of the proto-Earth and Theia in the same local disk environment to resolve this issue is difficult, as there is no guarantee that they would emerge with the same isotopic nature.
Instead, \cite{Canup_2012} demonstrated that a larger impactor-to-target mass ratio of $\phi \approx 0.4-0.5$ allows more mixing between the proto-Earth and Theia, with the impactor contributing equally to the planet and disk produced from the collision. Such a merger of a $\sim 0.6-0.7$ proto-Earth and a $\sim 0.3-0.4 M_\oplus$ Theia would still satisfy the angular momentum constraint of the Earth-Moon system \citep{cuk_2012}. 
A Theia mass of $\sim 0.4 M_\oplus$ has also been proposed by \cite{Desch_2019} to explain the preservation of ingassed hydrogen from Theia to the Moon.

Moreover, comparisons of $^3$He/$^{22}$Ne ratios between the upper degassed and lower preserved mantles suggest that an additional giant impact must have shaped the Earth's current internal structure. 
Giant impacts are capable of re-melting parts of the Earth's mantle, re-creating global magma oceans and enabling the formation of secondary atmospheres via mantle outgassing \citep{Elkins-Tanton_2012}.
Each outgassing/ingassing cycle increases $^3$He/$^{22}$Ne values in the mantle, since He is preferentially dissolved in basaltic melts compared to Ne \citep{Jambon_1986,Iacono-Marziano_2010}. Consequently, \cite{Tucker_2014} established that the upper mantle likely underwent at least two phases of melting from impacts, raising its initial $^3{\rm He}/^{22}{\rm Ne}$ ratio from the preserved deep mantle signature of $^3{\rm He}/^{22}{\rm Ne}\sim 2.3-3$ to the present upper mantle value of $\geq 10$.

\subsubsection{A cohesive Earth formation pathway}
\label{subsubsec:cohesive_formation_pathway}

In light of the current constraints on the final assembly of the Earth, we propose in Figure \ref{fig:favored_formation_scenario_impacts} a cohesive pathway to assemble the Earth, beginning with the emergence of a set of three $\sim 0.3-0.4 M_\oplus$ embryos in the late stages of the solar nebula. As discussed in Section \ref{subsubsec:late_stage_formation}, there is limited flexibility to envision alternative initial configurations of Earth embryos due to the tight constraints on the critical embryo mass of $\sim 0.3-0.4 M_\oplus$ in the later phases of the solar nebula.

In this scenario, the three Earth ancestors complete their assembly in the same local ($\sim$1 au) environment of the depleted ($f_{\rm dep} \sim10^{-4}-10^{-1}$) gas disk. At this time, they each rapidly accrete a primordial envelope and deliver the required concentration of primordial $^{22}$Ne of $c_{\rm ^{22}Ne,p} \approx (1.08 \pm 0.60) \times 10^{-13} \mathrm{mol/g}$ as demonstrated by Figure \ref{subfig:Ne_0.3}.
 In the presence of an early, gas-rich background nebula, gas damping efficiently circularizes orbits, hindering further embryo growth via giant impacts until disk dispersal.
As the nebula dissipates, the damping of embryos' eccentricities $e$ is removed, enhancing the likelihood of their orbits crossing and triggering giant impacts \citep{Papaloizou_2000}.

We therefore propose the existence of an intermediary phase (third stage depicted in Figure \ref{fig:favored_formation_scenario_impacts}), following nebular dispersal ($t \sim 1-5\mathrm{Myr}$) but prior to the Moon-forming impact ($t \sim 30-100\mathrm{Myr}$), during which a young $\sim 0.6-0.7 M_\oplus$ proto-Earth coexist in the vicinity of the remaining $\sim 0.3-0.4 M_\oplus$ embryo. In this context, the latter would correspond to the hypothesized ancient planet Theia, with a comparable size to the proto-Earth, as allowed by current terrestrial and lunar chemical constraints (Section \ref{subsubsec:impact_constraints}).
We favor a local formation of Theia as it is expected to have a similar composition to the proto-Earth \citep{Mastrobuono_Battisti_2015,Dauphas_2017}.
Subsequently, the final Earth-Moon system would be reached through the Moon-forming impact (blue arrow in Figure \ref{fig:favored_formation_scenario_impacts}) between the proto-Earth and Theia.

Quantitatively, the feasibility of the proposed scenario of Figure \ref{fig:favored_formation_scenario_impacts} can be established by comparing the orbit-crossing timescale $t_X$ of the embryos of mass $m$ with their eccentricity damping timescales $t_{\rm damp} \equiv \frac{e}{\dot{e}}$.
\cite{Papaloizou_2000} established a criterion of $t_X<10t_{\rm damp}$ to assess whether a given protoplanetary disk would favor embryo mergers, based on their $N$-body dynamical simulations of agglomeration of protoplanetary cores.
As derived by \cite{Tanaka_2004}, the eccentricity damping timescale can be expressed as

\begin{equation}
    \label{eq:t_damp}
    t_{\rm damp} \sim \frac{c_{\rm s}^3}{G^2m\rho_{\rm disk}} \sim 816 \mathrm{yr} \frac{M_\oplus}{m} \frac{1}{f_{\rm dep}},
\end{equation}
where $c_s$ is evaluated in the MMSN disk at 1 au using Equations (\ref{eq:T_MMSN}) and (\ref{eq:c_s}) and $\rho_{\rm disk}$ with Equation (\ref{eq:rho_MMSN}).\footnote{For simplicity, the velocity of embryos relative to the gas is assumed to be $\sim c_s$, dominated by their eccentric and headwind motion, which is subsonic. This approximation is valid for embryos with eccentricities $e \lesssim c_s/v_K \sim 1/30$, where $c_s\sim 1{\ \rm km/s}$ (Equation \ref{eq:c_s}) and the Keplerian velocity $v_K \sim 30{\rm \  km/s}$ is evaluated at 1 au.}
Comparatively, \cite{Zhou_2007} developed an empirical formula for the orbit crossing timescale of the doubling of embryos of mass $m \rightarrow 2m$,

\begin{equation}
    \label{eq:t_X}
    t_X(m,e,k) = 10^{A(m,e,k)+B(m,e,k) \log_{10}(k/2.3)} \mathrm{yr}.
\end{equation}
Here, $k\equiv\Delta a/R_{\rm mH}$ is the orbital separations $\Delta a$ of the objects in units of their mutual Hill radii

\begin{equation}
    \label{eq:R_mH}
    R_{\rm mH} = \Big( \frac{2m}{3M_\odot} \Big)^{1/3} a = 0.0126 \Big(\frac{m}{M_\oplus} \Big)^{1/3}a.
\end{equation}
The parameters

\begin{equation}
    \label{eq:t_X_A}
    A = -2 + e/h - 0.27 \log_{10} \Big( \frac{m}{M_\odot} \Big)
\end{equation}
and 
\begin{equation}
    \label{eq:t_X_B}
    B = 18.7 - 16.8e/h + (1.1 - 1.2e/h) \log_{10} \Big( \frac{m}{M_\odot} \Big)
\end{equation}
are fitted constants, where

\begin{equation}
    \label{eq:t_X_h}
    h = \frac{k}{2} \Big( \frac{2m}{3M_\odot} \Big)^{1/3}.
\end{equation}

\begin{figure}[h]
    \centering
    \includegraphics[width=\linewidth]{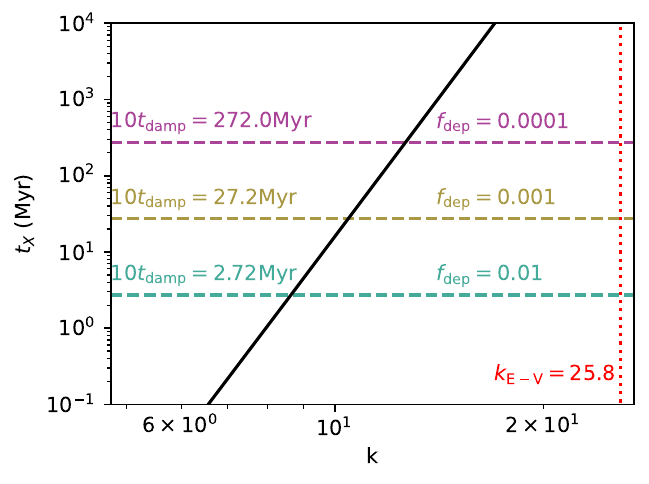}
    \caption{Orbit crossing time $t_X$ of $0.3M_\oplus$ embryos assembled at $a\sim 1$ au with eccentricity $e=0.001$, as a function of orbital spacing $k$. The eccentricity damping timescale $t_{\rm damp}$ of the embryos in a gaseous nebula depleted by factors $f_{\rm dep}=[10^{-2},10^{-3},10^{-4}]$ are displayed with cyan, yellow and magenta dashed horizontal lines, respectively. The damping times are multiplied by 10 to assess the merger criterion $t_X < 10 t_{\rm damp}$ of \cite{Papaloizou_2000}. As a reference, the orbital spacing $k_{\rm E-V}$ between the Earth and its nearest neighbor Venus is indicated with a red dotted vertical line.}
    \label{fig:t_X_vs_k}
\end{figure}

In Figure \ref{fig:t_X_vs_k}, we compute $t_X$ for a range of possible orbital separations between $0.3M_\oplus$ embryos, and compare it with $t_{\rm damp}$ in the depleted gas nebula. We fix $e=0.001$, as embryo eccentricities are expected to be small ($e \lesssim 0.01$) for such high mass embryos in the later phase of the life of protoplanetary disks \citep{Ida_2004}. 
Consistently with the $M_{\rm rock}\sim 0.3-0.4 M_\oplus$ solution of Figure \ref{fig:Ne_dissolved} with $10^{-4} \lesssim f_{\rm dep}\lesssim 10^{-1}$,
we find that a ``proto-Earth doubling" merger of two $0.3-0.4 M_\oplus$ embryos (denoted by an orange arrow in Figure \ref{fig:favored_formation_scenario_impacts}) would have been facilitated (i.e. $t_X<10t_{\rm damp}$) by the dispersal of the gas disk ($f_{\rm dep}\lesssim 10^{-3}$), with the orbits of embryos with $k \lesssim 10$ crossing within 30 Myr. 
Furthermore, eccentricity damping halts the mergers of $\sim 0.3-0.4 M_\oplus$ embryos in $f_{\rm dep} \gtrsim 10^{-2}$ disk for spacings $k \gtrsim 10$, explaining why the initial oligarchic growth of embryos (denoted with a gray arrow in Figure \ref{fig:favored_formation_scenario_impacts}) would have favored embryo masses of $\sim 0.3-0.4 M_\oplus$ in the nebula.
It is therefore reasonable to consider a first proto-Earth merger at $a\sim 1{\rm au}$ of Earth embryos before the Moon-forming impact at $t\sim30-100\mathrm{Myr}$, with an initial mean spacing between the embryos of $k\sim 10$, corresponding to orbital separations of $\Delta a \sim 0.08 \mathrm{\ au}$.

We note that an initial embryo spacing of $k\sim10$ for the entire inner solar system would be inconsistent with the current Earth-Mars spacing of $k_{\rm E-M} = 39.9$, but would marginally agree with the spacing of $k_{\rm E-V} = 25.8$ between the Earth and Venus.
Indeed, each global embryo doubling event ($m\rightarrow2m$) increases the average value of $k$ in a system by a factor of $2^{2/3}\approx1.6$ (by definition of $k$ and Equation \ref{eq:R_mH}). As such, a set of Venus, Earth and Mars embryos emerging from early assembly in the nebula with mean spacings $k\sim 10$ would lead to average spacings of $k\sim 25$ after two doubling events. The discrepancy with the Earth-Mars spacing may be a result of Mars not participating in later growth phases, consistent with the evidence of Mars being a stranded planetary embryo formed early in the nebula \citep{Dauphas_2011}.

The formation pathway of Figure \ref{fig:favored_formation_scenario_impacts} implicitly assumes that primordial volatiles stored in the rocky interior of embryos are preserved during these two collisions. Consequently, the primordial $^{22}$Ne concentration of the rocky interior of each embryo corresponds to that of the bulk mantle of the Earth, after correcting for mantle outgassing. 
Under certain impact conditions, it may be possible to strip only the upper mantle and preserve the deep mantle, as proposed by \cite{Tucker_2014}. 
If not, the fate of ejected volatiles in catastrophic collisions is unclear, and we argue that the ejecta would nonetheless be recaptured by the merged body and reincorporated into the interior of the merged protoplanet.
We will assess this statement by first computing whether the ejected material would be able to complete a full orbit, forming a torus along the path of the merged body, before being blown away by the solar wind.
We will also investigate whether this torus of ejected gas diffuses into the interplanetary medium before its re-accretion.

First, the dynamical pressure of the solar wind exerts an outward acceleration of

\begin{equation}
    a_{\rm dyn} \sim \frac{2\pi a \delta l \rho_* v_*^2}{M_{\rm ejected}},
\end{equation}
where $a$ is the orbital distance of 1 au, $\delta l$ the width of the torus, $\rho_*$ the volume mass density of solar wind, $v_*$ its velocity, and $M_{\rm ejected}$ the total mass ejected by the impact. Here, the dynamical pressure of the solar wind $P_{\rm SW} \sim \rho_*v_*^2$ is estimated using average values at 1 au of $v_*\approx 376 \mathrm{\ km/s}$ and $\rho_*\approx 6 \ m_H\mathrm{cm^{-3}}$ \citep[taken for the slow wind component in Table 4 of][neglecting alpha particles and heavier ions for simplicity]{Larrodera_2020}. 
We set the width of the torus to $\delta l \sim v_{\rm esc}/\Omega$, where $v_{\rm esc} \equiv \sqrt{\frac{2GM_{\rm rock}}{R_{\rm rock}}} \sim 11.2 {\rm \ km/s} \Big(\frac{M_{\rm rock}}{M_\oplus} \Big)^{1/3}$ is the surface escape velocity of the embryo and $\Omega$ is the orbital frequency at 1 au.
Consequently, over a single orbit of the ejected material, its radial velocity would have grown relative to the Keplerian local orbital velocity $v_K \equiv a\Omega$ by

\begin{equation}
    \label{eq:dv_vK}
    \frac{\Delta v}{v_K} \sim \frac{2\pi }{ \Omega} \frac{a_{\rm dyn}}{v_K} \sim 1.3 \times 10^{-8} \Big( \frac{M_{\rm rock}}{M_{\rm ejected}} \Big) \Big( \frac{M_\oplus}{M_{\rm rock}}\Big)^{2/3}.
\end{equation}

Second, the torus of ejected gas with width $\delta l$ will diffuse on a timescale of 

\begin{equation}
    \label{eq:t_diff_def}
    t_{\rm diff} \gtrsim \frac{\delta l^2}{\nu},
\end{equation}
where $\nu$ is the kinematic viscosity of the gas. As the viscosity of nebular gas is uncertain  \citep{Rafikov_2017}, we estimate with the molecular viscosity $\nu_{\rm mol} \sim v_{\rm th} \lambda /3$, where $v_{\rm th}$ and $\lambda \equiv 1/(\sqrt{2} n \sigma_{\rm mol})$ are the mean molecular speed and mean free path of the ejected medium, respectively, the latter set by the number density $n $ of molecules in the torus and the molecular cross-section $\sigma_{\rm mol}$ \citep{BirdStewartLightfoot2007}.
We express Equation (\ref{eq:t_diff_def}) as a lower limit on $t_{\rm diff}$ because full diffusion would occur over the entire orbital distance $a$, so that $t_{\rm diff}$ can also be estimated as $\sim a^2/\nu$.
We expect the random velocity of ejected particles to be comparable to the escape velocity of the merged embryos and therefore set $v_{\rm th} \sim v_{\rm esc}$. 
For simplicity, we further assume that $n$ is constant in space, so that the ejected mass $M_{\rm ejected}$ of primordial gas with mean molecular weight $\mu_{\rm disk} \approx 2.37$ (ignoring compositional changes from disk values) is uniformly distributed over the volume of the torus $2 \pi^2 a \delta l^2$. 
We adopt the cross section of H$_2$, that is $\sigma_{\rm mol} \sim \sigma_{\rm H_2} \approx 0.2 {\rm nm^2}$  \citep[see reaction 46 of ][]{TABATA_2000}, representative of a H-dominated gas in a molecular regime.
Under these assumptions, the timescale of diffusive loss is given by

\begin{equation}
    \label{eq:t_diff_result}
    t_{\rm diff} \gtrsim \frac{3}{\sqrt{2}\pi^2} \frac{\sigma_{\rm mol}}{a v_{\rm esc}} \frac{M_{\rm ejected}}{\mu_{\rm disk}m_H},
\end{equation}
or 

\begin{equation}
    \label{eq:t_diff_scaling}
    t_{\rm diff} \gtrsim 1.65 \times 10^{9} {\rm yr} \Big( \frac{M_{\rm ejected}}{M_{\rm rock} }\Big) \Big(\frac{M_{\rm rock}}{M_\oplus} \Big)^{2/3}.
\end{equation}

For a proto-Earth doubling consisting of two $\sim 0.3-0.4 M_\oplus$ embryos merging into a $M_{\rm rock} \sim 0.6-0.7 M_\oplus$ proto-Earth, we can fix as a lower bound for $M_{\rm ejected}$ the envelope mass $M_{\rm env} \approx 2\times 10^{-6} M_{\rm rock}$ (see the lower panel of Figure \ref{fig:example_envelope}) required to explain the deep mantle primordial Ne reservoir.
Using $M_{\rm rock} =0.65 M_\oplus$ for a representative case, Equations (\ref{eq:dv_vK}) and (\ref{eq:t_diff_scaling}) give $\Delta v/v_K \sim 0.0089 \ll 1$ and $t_{\rm diff} \gtrsim 2500 {\rm \ yr}$.
As such, primordial volatiles are unlikely to escape the orbit prior to their recapture by the formed protoplanet, instead  being rapidly reaccreted and dissolved into its rocky interior, on timescales comparable to those shown in Figure \ref{fig:Ne_dissolved}. 
This calculation also shows that the concentration of nebular $^{22}$Ne dissolved into embryos could not have initially exceeded our reference value of $c_{\rm ^{22}Ne,p}= (1.08 \pm 0.60) \times 10^{-13} \mathrm{\ mol/g}$, reinforcing our tight upper constraint on the critical embryo mass of $\sim0.3-0.4M_\oplus$ found in Section \ref{sec:results}. We acknowledge that these estimates are crude and will require future investigation.

\section{ Noble gas budget of the deep mantle}
\label{sec:Noble_gas_budget}

In this section, we discuss the current constraints on the origin of the deep-mantle budgets of He, Ne, Ar, Kr, and Xe, and whether our dissolution calculation, which assumes a fully nebular Ne budget, aligns well with the broader understanding of these reservoirs.
Along with the measurements of  $c_{\rm ^{22}Ne,DM}$ used as a reference in Section \ref{sec:results}, \cite{Marty_2012} (see their Table 1) reported the concentrations of the non-radiogenic $^3$He,\footnote{Although not directly reported by \cite{Marty_2012}, the bulk mantle concentration of $^3$He can be recovered from their measurements \citep[see Section 3.3 of ][]{Marty_2012} of $c_{\rm ^{14}N} = (9.0 \pm 4.6) \times 10^{-8} \mathrm{mol/g}$,  ${\rm N/}^{40}{\rm Ar}=160\pm40$, $^4 {\rm He/}^{40}{\rm Ar}=1.9\pm0.5$ and the helium isotopic ratios of $^3 \mathrm{He/}^4\mathrm{He}\sim 5.18 \times 10^{-5} $ measured in Icelandic OIB samples \citep{Hilton_1999}.} $^{36}$Ar, $^{84}$Kr, and $^{130}$Xe isotopes trapped in mineral lattice samples of the bulk (i.e. non-degassed) mantle, which we summarize in Table \ref{table:c_mantle}. 
The existence of a preserved nebular Ne reservoir in the mantle as supported by the analysis of \cite{Williams_2019} implies that other components of the proto-atmosphere were similarly incorporated into the rocky interior during the chemical equilibration of the mantle-envelope boundary.

\begin{table}[h]
\centering
\begin{tabular}{|c|c|}

\hline
${\rm ^AX}$ & $c_{\rm ^AX,DM}$ (mol/g)\\
\hline
$^3$He & $(5.7 \pm 3.6)\times 10^{-14}$\\
\hline
$^{22}$Ne & $(5.8 \pm 3.2)\times 10^{-15}$ \\
\hline
$^{36}$Ar & $(7.8 \pm 4.3)\times 10^{-14}$  \\
\hline
$^{84}$Kr & $(1.9 \pm 1.1)\times 10^{-15}$ \\
\hline
$^{130}$Xe & $(2.60 \pm 0.68)\times 10^{-17}$ \\
\hline
\end{tabular}

\caption{Noble gas concentrations $c_{\rm ^Ax,DM}$ of non-radiogenic isotopes (${\rm ^AX}$) of He, Ne, Ar, Kr and Xe reported by \cite{Marty_2012} for the Earth's bulk (i.e. non-degassed) mantle.}
\label{table:c_mantle}
\end{table}

\begin{figure}[t]
    \includegraphics[width=0.45\textwidth]{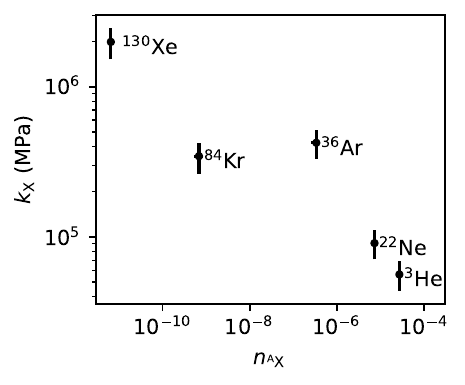}

    \caption{Dissolution parameters of 5 non-radiogenic isotopes (${\rm ^AX}$) of nebular noble gases ($^{3}$He, $^{22}$Ne,  $^{36}$Ar, $^{84}$Kr, and $^{130}$Xe) into basaltic melt. The abundances of each isotopes $n_{\rm ^AX}$ are taken from the analysis of \citet{Asplund_2021}. We use the elemental values of Henry's constant $k_{\rm X}$ (See Equation \ref{eq:Henrys_law}) of \cite{Iacono-Marziano_2010} for He, Ne and Ar and the ones of \cite{Jambon_1986} for Kr and Xe, evaluated at the melting temperature of the mantle $T_{\rm melt} \approx 1800 \mathrm{K}$ (see Figure \ref{fig:adiabat_vs_solidus}).}
    \label{fig:dissolution_parameters}
\end{figure}

The relative concentrations of primordial volatiles dissolved in the mantle are regulated by their primordial abundances in the solar nebula and their solubility in basaltic melts.
Given the concentration $c_{\rm  ^{22}Ne,p}$ of dissolved primordial $^{22}$Ne into the mantle at surface pressure $P_0$, Henry's Law (Equation \ref{eq:Henrys_law}) can be used to compute the expected initial primordial concentration $c_{\rm ^AX,p}$ of an arbitrary species ${\rm ^AX}$:

\begin{equation}
    \label{eq:primordial_ratios}
    c_{\rm ^AX,p}  = \frac{k_{\rm Ne}}{k_{\rm X}} \frac{n_{\rm ^AX}}{ n_{\rm  ^{22}Ne}} c_{\rm  ^{22}Ne,p}.
\end{equation}
Here, $k_{\rm X}$ and $n_{\rm ^AX}$ are the solubility constant and elemental abundance of species ${\rm ^AX}$, following the same format as for $^{22}$Ne in Equation (\ref{eq:Henrys_law}). 
As discussed in Section \ref{subsec:dissolution}, the composition of the solar nebula has been constrained by spectroscopic analyses of the Sun's photosphere from \cite{Asplund_2021}, and the model of \cite{Iacono-Marziano_2010} can once again be used, this time to compute the solubility constants of He and Ar. For Kr and Xe, we resort to the solubility constants measurements in basaltic melts of \cite{Jambon_1986}. The constant $k_{\rm X}$, which in Sections \ref{sec:methods}-\ref{sec:results} was evaluated as a function of $T_0$ for X=Ne, is fixed in Equation (\ref{eq:primordial_ratios}) to its value at $T_{\rm melt} \approx 1800 \mathrm{K}$ for simplicity. The elemental abundances and solubility constants of $^3$He, $^{22}$Ne, $^{36}$Ar, $^{84}$Kr and $^{130}$Xe are presented in Figure \ref{fig:dissolution_parameters}, which shows that lighter noble gases are preferentially dissolved into basaltic melts.

\begin{figure}[t]
    \includegraphics[width=0.45\textwidth]{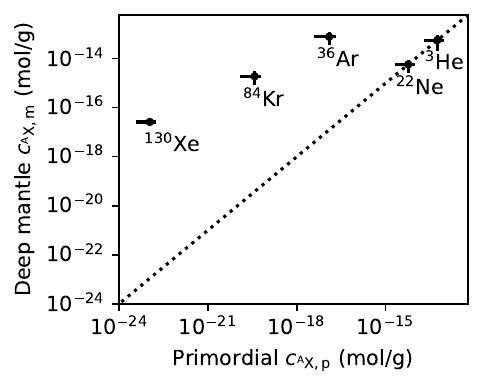}

    \caption{Comparison of the expected concentrations of primordial nebular gas dissolved in the mantle with the measured deep mantle concentrations measured by \cite{Marty_2012} in plume samples. A dotted one-to-one line centered on $^{22}$Ne is shown to illustrate how the $^{36}$Ar, $^{84}$Kr and $^{130}$Xe reservoirs cannot be purely explained by the capture of primordial volatiles as isotopic measurements argue is the case for Ne \citep{Williams_2019} and most likely He (see Section \ref{subsec:primordial_expectation}).}
    \label{fig:primordial_expectation}
\end{figure}

The remainder of this section explores the implications of Figure \ref{fig:primordial_expectation}, in which we compare the predictions of Equation (\ref{eq:primordial_ratios}) with the measured deep mantle concentrations of \cite{Marty_2012}.\footnote{Our calculations are anchored to Ne. To compare our predictions to present day reservoirs of each species, Equation (\ref{eq:primordial_ratios}) must  be corrected by an additional factor of $k_{\rm Ne}/k_{\rm X}$ for their preferential steady outgassing over geological timescales.}  In Section \ref{subsec:primordial_expectation}, we show that the solar-like He reservoir inferred from OIB samples is consistent with the nebular dissolution scenario. Section \ref{subsec:Ar_Kr_Xe_reservoirs} explains why similar conclusions cannot be drawn for Ar, Kr and Xe reservoirs, which appear to be mixtures of preserved early signatures of chondrite-like pristine source, contaminated by the subduction of gas from the Earth's secondary atmosphere. Finally, we summarize in Section \ref{subsec:consistent_mantle_budget} the current constraints on the noble gas content of the deep mantle, and present a simple calculation in which we explain the measurements of Table \ref{table:c_mantle} with a mixture of nebular, chondritic, and subducted atmospheric gas.

\subsection{Expected nebular deep mantle budget}
\label{subsec:primordial_expectation}

In Figure \ref{fig:primordial_expectation}, we report a $\approx 1\sigma$ agreement between the expected concentration of dissolved nebular ${\rm ^3He}$ of $(5.6 \pm 3.6) \times 10^{-14} \mathrm{g/mol}$ and its measured deep mantle budget of $(5.5 \pm 3.5) \times 10^{-14} \mathrm{g/mol}$. This agreement is consistent with a nebular source for the deep mantle He, but a SWI origin cannot be excluded from analyses of ${\rm ^3 He/^4He}$ ratios in OIB samples alone.

Indeed, He from the solar nebula and from SWI solids are characterized by ${\rm ^3 He/^4He}$ values of $\approx 259 R_{\rm A}$  and $(331.4\pm 6.4)R_{\rm A}$, respectively \citep{Heber_2012},\footnote{These associated values from \cite{Heber_2012} assume that isotopic compositions of the primordial nebula is consistent with the Sun's photosphere and the SWI solids with bulk solar wind measurements.} with $R_{\rm A} = 1.4 \times 10^{-6}$ being the atmospheric ${\rm ^3 He/^4He}$ value \citep{Sano_1988}.
Determining which source is most consistent with the initial He isotopic composition of the deep mantle is challenging, mainly because the $\alpha-$decay of heavy elements such as U and Th produces additional ${\rm ^4He}$ in the mantle, lowering the ${\rm ^3 He/^4He}$ ratio of the primitive non-radiogenic source \citep{Day_2005,Day_2015}.

Mantle plumes found in Hawaii \citep{Valbracht_1997}, Iceland \citep{Hilton_1999}, the Galápagos Islands \citep{Kurz_2009} and Samoa \citep{Jackson_2009} do exhibit high ${\rm ^3 He/^4He}\gtrsim 30 R_{\rm A}$ ratios \citep[see][their Table 2]{Day_2022} compared to the canonical MORB value of ${\rm ^3 He/^4He}=8\pm2 R_{\rm A}$ and the atmospheric value $R_A$, but relating these measurements to a precise source requires an understanding of how radiogenic He is produced in the Earth's silicate mantle \citep{Day_2022}.
As such, we limit ourselves to concluding from Figure \ref{fig:primordial_expectation} that our dissolution calculation of nebular Ne in the deep mantle as reported by \cite{Williams_2019} is consistent with the current knowledge of He from OIB samples.

\subsection{ Origins of heavy noble gases in the deep mantle}
\label{subsec:Ar_Kr_Xe_reservoirs}

Figure \ref{fig:primordial_expectation} displays an excess of $^{36}$Ar, $^{84}$Kr and $^{130}$Xe by factors of $\gtrsim 10^4$ compared to the nebular dissolution scenario. 
If nebular accretion cannot fully account for this excess in the deep mantle, what processes supply the heavy noble gas content?

First, carbonaceous chondrites (e.g. type CI) are volatile-rich primitive meteorites \citep{Kallemeyn_1981} and were incorporated into the Earth during its assembly \citep{Pepin_1991,Marty_2012,Halliday_2013}. They carry most of their noble gases in a carrier phase called the ``Q-component" \citep[originally named for “quintessence” by][]{Lewis_1975} with chemical and isotopic compositions distinct from solar values, enriched in heavy noble gases \citep{Ott_2002}. 

Second, the recycling of atmospheric volatiles via dissolution in the oceanic lithosphere and subsequent subduction in the Earth's interior is an efficient carrier of heavy noble gases \citep{Ballentine_2000,Holland_2006,Kendrick_2011,Mukhopadhyay_2019}. 
Although this contamination from the Earth's secondary atmosphere is not a viable source of light noble gases (He, Ne) due to their poor solubility in seawater \citep{KENDRICK_2018}, recent isotopic analysis of Galápagos and Iceland OIB samples along with realistic recycling calculations by \cite{Peron_2022} argued that significant fractions of Kr and Xe found in the deep mantle (in the concentrations shown in Table \ref{table:c_mantle}) are of atmospheric provenance.\footnote{Another species that behaves like a noble gas in silicate melts is nitrogen (N$_2$) \citep{Roskosz_2013}. In the bulk mantle, N$_2$ is sourced by Earth's secondary atmosphere \citep{Marty_2003}, with $c_{\rm ^{14}N, DM} = (9.0 \pm 4.6) \times 10^{-8} {\rm mol/g}$ \citep{Marty_2012}.
With comparable solubilities in silicates \citep{Roskosz_2006} and solar abundances \citep{Asplund_2021} to $^{22}$Ne, we expect from Equation (\ref{eq:primordial_ratios}) similar primitive deep mantle concentrations of $^{14}$N and $^{22}$Ne (i.e. $c_{\rm ^{14}N,p}\approx c_{\rm ^{22}Ne,p}\approx c_{\rm ^{22}Ne,DM} = (5.8 \pm 3.2)\times 10^{-15} {\rm mol/g}$).
Any primordial nitrogen signatures in the deep mantle are therefore overprinted by atmospheric contamination and we refrain from including this volatile species in our discussion.} 

For Kr, approximately half of the deep mantle reservoir is isotopically consistent with air \citep{Peron_2021}. Disentangling primitive Kr signatures from atmospheric contamination requires high-precision isotopic measurements \citep[e.g., see Figure 2 of][]{Halliday_2013}. As illustrated in Figure \ref{fig:primordial_expectation}, the non-atmospheric half of the Kr budget is unlikely to originate from the solar nebula, and \cite{Halliday_2013} concluded from their analysis of ${\rm ^{84}Kr/^{36}Ar}$ in the deep Earth that it can be explained by the incorporation of a primitive volatile source, isotopically similar to carbonaceous chondrites, during the Earth's (or its parent bodies') assembly, an idea originally suggested by \cite{Holland_2009}.
The relative ratio of ${\rm ^{84}Kr/^{22}Ne} = 2.7 \pm 3$ \citep[see Tables 2, 3, and 4 of][]{Ott_2002} in the Q-component of chondrites is much greater than the value of ${\rm ^{84}Kr/^{22}Ne} = (1.6 \pm 0.9) \times 10^{-5}$ expected by the nebular dissolution scenario (see Figure \ref{fig:primordial_expectation}).
Consequently, an initial Kr reservoir provided by such chondritic volatile source would account for the large excess of $^{84}$Kr in the deep mantle while contributing negligibly to its Ne budget.

Several studies of chondritic signatures in the atmosphere and upper mantle \citep[e.g., see ][ for a review]{Walker_2009} have linked the acquisition of chondrites to a late veneer \citep{Morgan_2001}. On the contrary, the same cannot be assumed for Kr signatures in the deep mantle \citep{Marty_2012}, which is largely isolated from late veneer contributions \citep[see Figure 4 of ][ for a schematic illustration of the evolution of the Earth's accretion of chondrites/planetary in its global core-mantle-atmosphere system]{Harper_1996}. From their measurements of Kr isotopes in Galapagos and Iceland plumes, \cite{Peron_2021} concluded that chondritic Kr must have been delivered in the early accretion phases of the Earth, before the lower mantle differentiated itself from the upper layers and preserved its initial signatures. In fact, the coexistence of nebular Ne and chondritic Kr in the deep mantle implies that the primitive chondritic volatiles were incorporated into the Earth in its earliest stages of accretion, in the presence of the solar nebula \citep{Broadley_2020}.

This conclusion can easily be reconciled with our favored formation scenario (Figure \ref{fig:favored_formation_scenario_impacts}), simply by assuming that a volatile-rich chondritic source was incorporated into $\sim 0.3-0.4 M_\oplus$ Earth embryos as they formed their rocky interior, prior or simultaneously to their accretion of a primordial envelope and the dispersal of the disk gas. 
It is believed that the bulk of Earth's building blocks were of composition similar to volatile-poor enstatite chondrite \citep{Dauphas_2017}. By comparing the volatile abundances of carbonaceous chondrites to that of the bulk Earth (atmosphere + interior), \cite{Marty_2012} find that a balance between ``dry" volatile-poor material and a small $\sim (2 \pm 1) {\rm mol\%}$ contribution from e.g. carbonaceous chondrites would explain the bulk Earth budget of C, H$_2$O, Ne, Ar, and Kr (see their Figure 7 and Table 1).

For Xe, the deep mantle reservoirs are even more affected by atmospheric contamination, accounting for more than 90$\%$ of the Xe total concentration \citep{Peron_2022} and almost entirely overprinting any primordial Xe signature.
It is still generally understood that the Xe content found in the mantle also constitute a mixture of subducted atmospheric gas \citep{Holland_2006} and a minor \citep[$\lesssim 10\%$ according to][]{Peron_2022} primitive chondritic component \citep{Broadley_2020}. 
Correcting for recycled atmospheric noble gases via subduction, \cite{Peron_2022} constrained a value of ${\rm ^{130}Xe/^{84}Kr} \lesssim 30 \times 10^{-4}$ for the primitive component of the Galápagos and Iceland deep mantle, less than the expectation of ${\rm ^{130}Xe/^{84}Kr} \sim 1370 \times 10^{-4}$ for carbonaceous chondrites by more than an order of magnitude (see their Tables 1 and 3).
Comparison of primitive Kr and Xe reservoirs consequently implies that a significant fraction of the expected chondritic Xe in the deep mantle is ``missing".\footnote{ In Earth's present-day atmosphere, non-radiogenic Ne, Ar and Kr exist in chondritic proportions while Xe is depleted by $\sim 10\%$ \citep[see Figure 6 of][]{Marty_2012}. This long-standing ``atmospheric Xe deficit" \citep[see ][ for a recent review]{Bekaert_2025} is separate from the deep mantle Xe deficit discussed here, the latter being even greater.} 

To explain the low Xe/Kr ratio of primitive deep mantle reservoirs, \cite{Peron_2022} proposed three non-mutually exclusive scenarios. First, this isotopic trend may have been directly inherited from Earth's building blocks. The primitive ``chondrite-like" source of heavy noble gases would have Kr and Xe isotopic compositions similar to chondrites but with Xe/Kr deficits of two orders of magnitude relative to chondrites, although \cite{Peron_2022} note that such a volatile source has not yet been identified.
Second, Xe/Kr values decrease during magma outgassing events due to the higher solubility of Kr in basaltic melt (see Figure \ref{fig:dissolution_parameters}). This scenario is unlikely since outgassing events would concurrently increase $^3$He/$^{22}$Ne ratios above primitive $\lesssim 4$ values which is not observed \citep{Tucker_2014}.

A third viable hypothesis is the preferential partitioning of Xe in the iron core of the Earth. As investigated by \cite{Wang_2022}, the early formation of the outer core via the segregation of liquid iron from magma oceans is expected to have entrained volatiles in the iron core. The relative concentration of each noble gases expected in primordial reservoirs within the liquid core would be set by their metal/silicate partition coefficients $D$. 
Unlike Ne, Xe is siderophile ($D_{\rm Xe} \sim 1-10$) for high core-mantle boundary temperatures \citep[See Figure 2 of][]{Wang_2022}.
Specifically, the preferential partitioning ratio of Xe compared to Kr is $D_{\rm Xe}/D_{\rm Kr}\approx 10.7\pm 7.2$ \citep[See Equations 18-19 of][]{Wang_2022}.\footnote{\cite{Wang_2022} report partition coefficients as functions of temperature, which for our study would have to be set to the time-dependent core-mantle boundary temperature of early $\sim0.3-0.4 M_\oplus$ Earth embryos. For simplicity, we evaluate $D_{\rm Xe}/D_{\rm Kr}$ at $5500{\rm \ K}$ which corresponds core-mantle temperature recovered in the $T_0 = 1800 {\rm K}$ interior profile of Figure \ref{fig:adiabat_vs_solidus}. We note that the core-mantle temperature used is an approximate extrapolation of Equations (18-19) of \cite{Wang_2022} justified by small ($\lesssim 50\%$) variations in $D_{\rm Xe}/D_{\rm Kr}$ over the temperature regime ($2300{\rm K} \leq T \leq 5000{\rm K}$) they studied.} The Xe ``paradox" in the deep mantle could therefore be solved with most of the missing primitive Xe reservoir being incorporated in the iron core during early formation much more so than Kr, as we argue in Section \ref{subsec:consistent_mantle_budget}.

Finally, the origins of Ar in the deep mantle are more uncertain, with mantle reservoirs heavily dominated by the radiogenic $^{40}$Ar isotope, produced by the decay of $^{40}$K in the rocky interior and then outgassed in the atmosphere \citep{Bender_2008}.
Recovering an initial component using the relative abundances of the non-radiogenic $^{36}$Ar and $^{38}$Ar within the resulting $^{40}$Ar dominated mixture remains a challenge, and will likely require further samples and high-precision isotopic analyses  \citep{Day_2022}. 
In fact, the measured ${\rm ^{38}Ar/^{36}Ar}$ ratios of MORB and OIB samples are consistent with both atmospheric and chondritic values \citep{Raquin_2009,Peron_2017}. Hence, any solar-like primordial signatures of Ar must have been obscured by the incorporation of chondrites and/or subducted atmospheric gas.

In their discussion of subducted recycled noble gases, \cite{Peron_2022} hypothesized fractions of recycled non-radiogenic Ar ranging from $14\%$ to $50\%$.
They proposed these values to reflect a continuous range between the primitive Ne reservoir which shows no evidence of significant atmospheric contamination \citep{Williams_2019} and the approximate half of the Kr budget which is not sourced by chondrites but by subduction of air, as inferred from Kr isotopic measurements \citep{Halliday_2013,Peron_2021}.
Other studies argue that the fraction of subducted atmospheric Ar should be similar to that of Xe, with estimated values of $\sim 93\%$ and even $100 \%$ estimated by \cite{Holland_2006} and \cite{Williams_2019}, respectively.
The current lack of constraints on non-radiogenic Ar makes it a particularly poor tracer of the initial inventory of the deep mantle.

To summarize, we argue that the deep mantle primitive reservoirs of Ar, Kr and Xe originate from the inclusion of chondritic volatiles into the rocky interior Earth embryos over the earliest stages of their solid assembly in the nebula, while He and Ne in the deep mantle are sourced by the accretion and dissolution of nebula gas.

\subsection{ Towards a self-consistent deep mantle noble gas budget}
\label{subsec:consistent_mantle_budget}

\begin{figure*}[t]
    \centering
    \includegraphics[width=\linewidth]{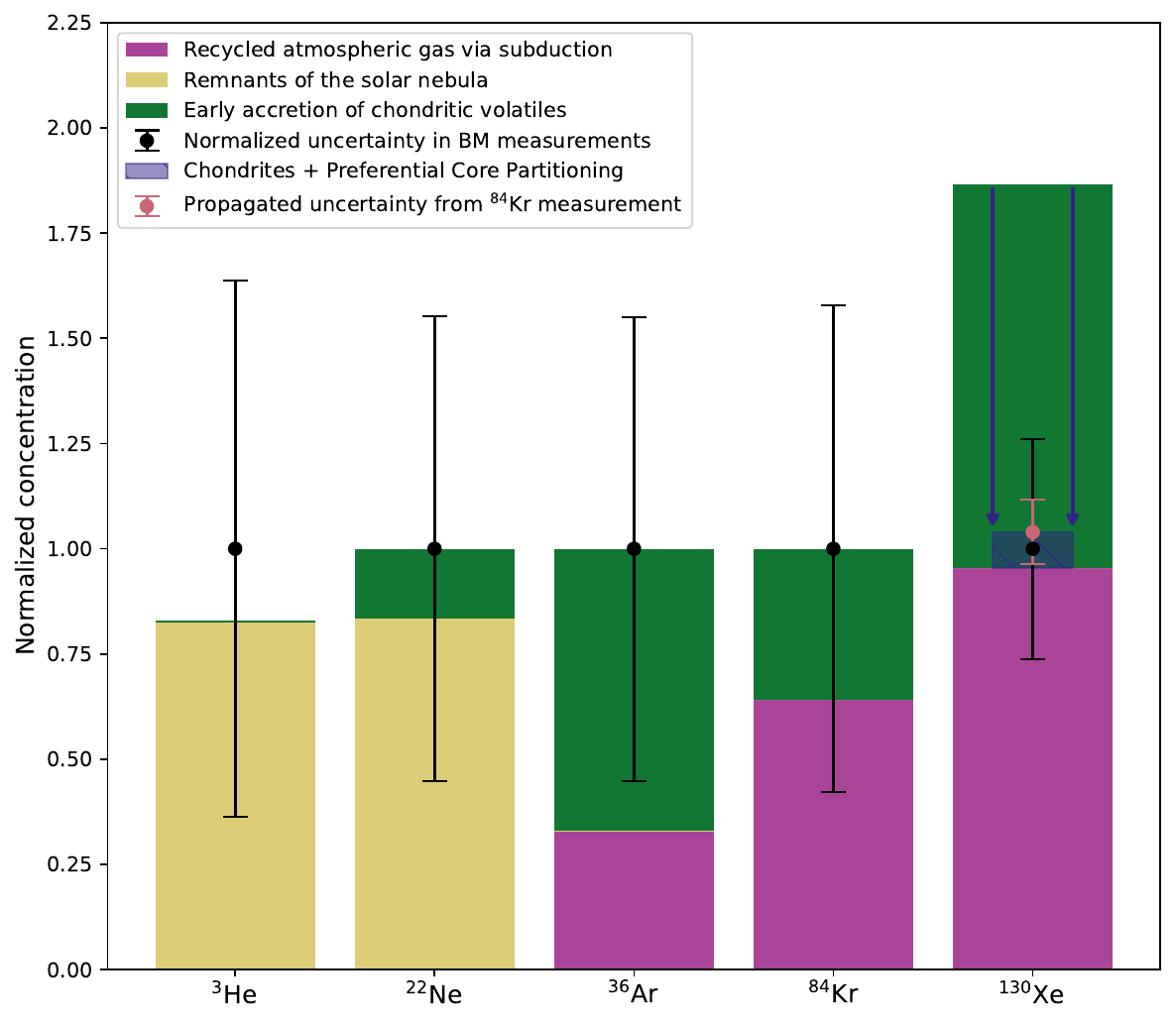}
    \caption{Estimates of the noble gas sourcing in the deep mantle reservoirs of $^3$He, $^{22}$Ne, $^{36}$Ar, $^{84}$Kr, and $^{130}$Xe. Contributions from dissolved primordial gas from the solar nebula, chondritic volatiles incorporated during early formation of rocky interiors and recycled gas subducted from the Earth's secondary atmosphere are represented in yellow, green and magenta, respectively. Concentrations are normalized to the measurements of noble gas concentrations in the bulk mantle (BM) of the Earth reported by \cite{Marty_2012}, with the uncertainty on those measurements shown in black. Due to the preferential partitioning of Xe into the core with respect to Kr \citep{Wang_2022}, the expected fraction of the initial $^{130}$Xe chondrite contribution is reduced and shown with a small inset blue bar, with the propagated uncertainty from the $^{84}$Kr measurement shown in pink.}
    \label{fig:Bar_chart}
\end{figure*}

In Figure \ref{fig:Bar_chart}, we summarize our discussion of noble gases in the deep mantle by presenting a simple estimation of the contributions of each source of the non-radiogenic $^3$He, $^{22}$Ne, $^{36}$Ar, $^{84}$Kr, and $^{130}$Xe. For simplicity, we use CI carbonaceous chondrites, rich in heavy noble gases, as a proxy for the primitive chondritic source discussed in Section \ref{subsec:Ar_Kr_Xe_reservoirs}, but note that other chondrite classes may explain the deep mantle signatures of heavy noble gases.
For comparison purposes, the contributions are normalized to the measured concentrations $c_{\rm ^AX,DM}$ of \cite{Marty_2012} presented in Table \ref{table:c_mantle} and can therefore be expressed as relative fractions (e.g., $f_{\rm ^{84}Kr,cc}\equiv \frac{ c_{\rm ^{84}Kr,cc}}{c_{\rm ^{84}Kr,DM}}$ for the fraction of the deep mantle $^{84}$Kr coming from carbonaceous chondrites). The figure is created as follows:

\begin{enumerate}
    \item We first fix the fractions $f_{\rm ^AX,atm}$ of each species ${\rm ^AX}$ accounted for by the subduction of atmospheric gas to the deep mantle. For Kr and Xe, we adopt the values reported by \cite{Peron_2022} in Icelandic plume sources of $f_{\rm ^{84}Kr,atm} = 64 \pm 8 \%$ and $f_{\rm ^{132}Xe,atm} = 95.4_{-5.5}^{+4.6} \%$, respectively.\footnote{Those measurements should be treated as upper bounds, as the analogous Galápagos OIB samples exhibit lower fractions of subducted atmospheric material of $f_{\rm ^{84}Kr,atm} = 48 \pm 8 \%$ and $f_{\rm ^{132}Xe,atm} = 93.4 \pm 4.6 \%$.}  
    As mentioned in Section \ref{subsec:primordial_expectation}, $^3$He and $^{22}$Ne in the deep mantle are expected to be primordial with little to no contamination from the Earth's secondary atmosphere, and we consequently assume for simplicity that these reservoirs are purely primitive (i.e. $f_{\rm ^{3}He,atm} \approx f_{\rm ^{22}Ne,atm} \approx 0$).  
    Because the exact balance of the $^{36}$Ar budget between a primitive chondritic source and subducted atmospheric gas remains uncertain, we fix a value of $f_{\rm ^{36}Ar,atm} \approx 33 \%$ to directly match the reported concentration of $^{36}$Ar by \cite{Marty_2012}, which is chosen after constraining the contribution from chondrites in the next step.
    This choice is quite arbitrary as the large uncertainty in the $^{36}$Ar measurement of \cite{Marty_2012} allows fractions ranging from 0\% to 73\%.

    \item Next, we anchor $f_{\rm ^{84}Kr,cc}$ to $100\% - f_{\rm ^{84}Kr,atm} \approx 36 \%$, attributing the remaining reservoir of $^{84}$Kr to chondritic volatiles. Based on the relative abundances $\Big(\frac{{\rm ^{A}X}}{\rm ^{84}Kr} \Big)_{\rm cc}$ of each noble gas ${\rm^AX}$ with respect to $^{84}$Kr in the primary Q component of chondrites \citep[computed from Tables 2, 3, and 5 of][]{Ott_2002}, we calculate the expected fraction of the other noble gas reservoirs that would consequently originate from chondrites, as given by
    \begin{equation}
    \label{eq:missing_Xe_correction}
        f_{\rm ^{A}X,cc} \approx f_{\rm ^{84}Kr,cc}\frac{ c_{\rm ^{84}Kr,DM}}{c_{\rm ^{A}X,DM}}\Big(\frac{{\rm ^{A}X}}{\rm ^{84}Kr} \Big)_{\rm cc} { \frac{k_{\rm Kr}}{k_{\rm X}}}. 
    \end{equation}
    The factor of $k_{\rm Kr}/k_{\rm X}$ accounts for different outgassing rates of the primitive chondritic reservoirs over geological timescales.
    Equation (\ref{eq:missing_Xe_correction}) gives fractions of $f_{\rm ^{3}He,cc} = 0.51 \pm 0.35\%$, $f_{\rm ^{22}Ne,cc} = 17 \pm 11 \%$, $f_{\rm ^{36}Ar,cc} = 67 \pm 41 \%$ and $f_{\rm ^{130}Xe,cc} = 91 \pm 54 \%$.
    The value of $f_{\rm ^{130}Xe,cc}$ is  remarkably high when compared to the atmospheric component ($f_{\rm ^{132}Xe,atm} = 95.4_{-5.5}^{+4.6} \%$),  emphasizing the ``paradoxal" Xe deficit when compared to the minor $\lesssim 10 \%$ chondritic component inferred by \cite{Peron_2022}.
    As argued in Section \ref{subsec:Ar_Kr_Xe_reservoirs}, Xe is more likely to partition into the core than Kr, which can reconcile the mantle Xe deficit problem. Using the relative partitioning factor of $D_{\rm Xe}/D_{\rm Kr}\approx 10.7 \pm 7.2$ \citep{Wang_2022}, we correct the value of $f_{\rm ^{130} Xe,CI}$ given by Equation (\ref{eq:missing_Xe_correction}) to $f_{\rm ^{130}Xe,cc} = (91 \pm 54 \%) /(10.7 \pm 7.2) =  8.5  \pm 7.6 \%$. The implication of this correction is that the vast majority of Xe supplied by chondrites is stored in the core during early assembly of embryos.
    \item Finally, we assume a $^{22}$Ne budget primarily dominated by remnants of the primordial solar nebula with $f_{\rm ^{84}Ne,p}=83 \%$, taking into account the small $\sim 17\%$ chondritic $^{22}$Ne contribution retrieved in the previous step. Using Equation (\ref{eq:primordial_ratios}) for normalized contributions then gives $f_{\rm ^{3}He,p} = 83 \pm 53\%$, $f_{\rm ^{22}Ar,p} = 0.013 \pm 0.009  \%$, $f_{\rm ^{36}Kr,p} = 0.0017 \pm 0.0011 \%$ and $f_{\rm ^{130}Xe,p} =  0.000035 \pm 0.000023 \%$.
\end{enumerate}

On the whole, Figure \ref{fig:Bar_chart} shows that the noble gas budgets of the deep mantle can be reconciled by a combination of dissolved nebular gas, incorporation of carbonaceous chondrites, and subducted atmospheric material,
without invoking the inclusion of SWI material.  
We conclude that light noble gases in the deep mantle reflect the gas dissolution of early Earth embryos while heavy noble gases complementarily probe the Earth's solid accretion history.

Assuming chondrites volatiles supplied the primitive Ar, Kr and Xe reservoirs, a ($\sim 17\%$) imprint of this primitive source may be present within the otherwise nebular Ne content. This conclusion is however subject to compositional variations between the chondritic primitive volatiles and CI chondrites. Such a contribution would be consistent (to $\approx 1 \sigma$) with the measurements of \cite{Williams_2019}, which allow a maximum of $10-15\%$ of the Ne reservoir to be of either atmospheric or chondritic origin \citep{Peron_2022}.

Lastly, Figure \ref{fig:Bar_chart} also demonstrates that the preferential partitioning of Xe in the Earth's iron core is a viable solution to the missing Xe problem in the deep mantle, reconciling to $\approx 1\sigma$ the measured concentration of Kr and Xe of \cite{Marty_2012}. A caveat to this last conclusion is that we have ignored the similar preferential partitioning of He, Ne and Ar. Their incorporation rate varies much more with temperature \citep[see Figure 2b of ][]{Wang_2022} and is harder to approximate without a complete model of the interior structure of planetary embryos.

We caution against drawing firm conclusions from this simple attempt at reconciling the measurements of He, Ne, Ar, Kr, and Xe in OIB samples, especially in light of the large uncertainties on deep mantle volatile measurements \citep{Marty_2012,Halliday_2013}. 
More precise measurements and detailed models of planetary interior dynamics beyond the scope of this work are needed to further constrain a self-consistent narrative for the origins of all five noble gas reservoirs (He, Ne, Ar, Kr, and Xe).

\section{ Primordial dissolution of hydrogen}
\label{sec:H_dissolution}

The capture of primordial atmospheres atop the molten surface of Earth embryos delivers H$_2$ to their interior, consistent with the dissolution of noble gases. The chemical interactions of nebular H$_2$ with silicate species such as MgO, SiO$_2$, MgSiO$_3$, FeO, MgSiO$_3$ prevents us from using H as a probe of early accretion processes, but its delivery to the rocky interior of Earth embryos may explain several of the Earth's key features \citep{Young_2023}. 

First, once incorporated into the mantle, hydrogen can be converted to water through its interactions with silicates.
As such, a nebular origin is a plausible explanation of the Earth's water content \citep{Genda_2008}. 
Second, the Earth's outer \citep{Birch_1964} and inner \citep{Jephcoat_1987,Stixrude_1997} cores are significantly less dense than pure Fe, by $\approx 10\%$ \citep{Jeanloz_1979} and $\sim 3-6\%$ \citep{Anderson_1994}, respectively. This core density deficit implies the presence of light volatiles in the metal core, with C, H, O, S and Si as the most likely candidates \citep[see ][ for a review]{Hirose_2021}. Once H$_2$ is dissolved into the mantle, \cite{Okuchi_1997} argued that $\gtrsim95\%$ of the freshly formed water should have reacted with iron, partitioning H to the metal core and possibly accounting for up to 60\% of the core density deficit.
Lastly, following nebular dispersal, H$_2$ degasses from the mantle and is subsequently lost by hydrodynamic escape.
Because H$_2$ acts as a reducing agent, \cite{Sharp_2017} inferred that its removal increases the Earth's bulk oxidation state, consistent with the mantle being highly oxidized. 

Using a thermodynamical model of the chemical interactions of primordial atmospheres and magma oceans, \cite{Young_2023} showed that these three geochemical constraints can consistently be explained by the chemical equilibrium of volatiles (O, H, C)
at the interface of the atmosphere and the magma.
Starting with $0.2-0.5M_\oplus$ Earth embryos of similar structure (see their Figure 3) to that of Figure \ref{fig:structure_diagram}, they found that initial H$_2$ primordial envelopes making up 0.2\% of the total embryo mass would approximately produce one terrestrial ocean and a metal density deficit of $8\%$, with final oxidation states consistent with the bulk silicate Earth. 
In light of the results presented in Section \ref{sec:results}, a natural question to ask is whether the favored setting of \cite{Young_2023} also reproduces the abundance of nebular $^{22}$Ne reported by \cite{Williams_2019} in the deep mantle. 

\begin{figure}[t]
    \includegraphics[width=0.45\textwidth]{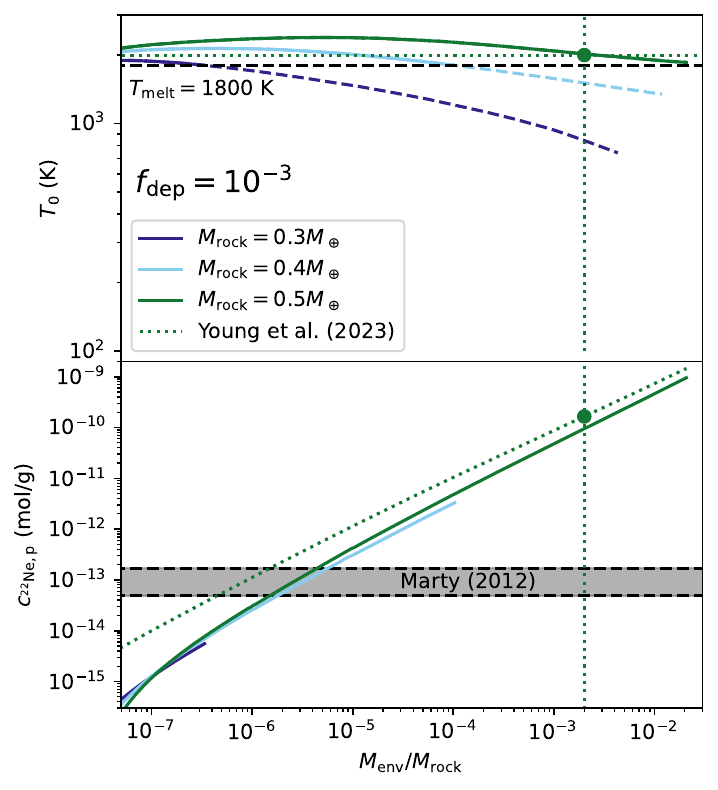}

    \caption{Surface temperature $T_0$ (upper panel) between the basaltic mantle of Earth embryos and primordial gas envelopes, along with the resulting concentration of primordial neon $c_{\rm ^{22}Ne,p}$ dissolved in the mantle (lower panel), both as functions of $M_{\rm env}/M_{\rm rock}$. 
    We compare the input atmospheric parameters and the resulting dissolved neon content (given by Equation \ref{eq:Henrys_law}) of the solution reported by the thermodynamic model of \cite{Young_2023} (dots and dotted curves)  with our results for rocky interior of masses $M_{\rm rock} =[0.3,0.4,0.5]M_\oplus$ in blue, cyan and green, respectively. The embryos are embedded in a minimum-mass solar nebula of gas density depleted by a factor $f_{\rm dep}=10^{-3}$ (see Equation \ref{eq:rho_MMSN}).
    Volatiles are only dissolved in the presence of surface magma oceans (solid curves), that is when $T_0$ is greater than the melting temperature of the silicate mantle $T_{\rm melt} \approx 1800{\rm K}$ (Section \ref{subsec:limations}), displayed with a black dashed curve. Once the surface of the mantle solidifies (colored dashed curves), atmospheric volatiles no longer dissolve in the solid surface of the rocky interior. We only show envelope masses corresponding to $t_{\rm acc} \lesssim 10 {\rm \ Myr}$, before envelope growth is cut short by the dispersal of the disk. The target concentration of primordial neon required to explain the measurements of \cite{Marty_2012} is shown with a gray shaded band.} 
    \label{fig:comparison_young+23}
\end{figure}

In Figure \ref{fig:comparison_young+23}, we compare the initial conditions (prior to the exchange of H with the mantle) of the atmosphere used by \cite{Young_2023} to obtain their solution with our gas accretion calculation. We only present our results for $f_{\rm dep} = 10^{-3}$ since they are in agreement with the input parameters of $M_{\rm env}/M_{\rm core} \approx M_{\rm env}/M_{\rm p} = 0.2\%$, $T_0 = 2350{\rm K}$, and 

\begin{equation}
    \label{eq:P0_Young}
    \Big( \frac{P_0}{1 {\rm \ MPa}} \Big) = 1.2 \times 10^5 \frac{M_{\rm env}}{M_p} \Big(\frac{M_p}{M_\oplus}\Big)^{2/3},
\end{equation}
used by \cite{Young_2023} for their $0.5 M_\oplus$ solution.\footnote{We note that the values of $P_0$ obtained from our accretion model are smaller (by $\lesssim 10\%$) than the approximate scaling of Equation (\ref{eq:P0_Young}), which results in a negligible offset in the dissolved concentrations reported in the lower panel of Figure \ref{fig:comparison_young+23}. We also note that solar-metallicity gas with $\mu_{\rm disk} \approx 2.37$ is considered in this work while \cite{Young_2023} initialized their model with a $99.9{\rm mol}\%$ H envelope of $\mu = 2.04$.}
Even though they only present their results for $0.5 M_\oplus$ embryos, they state that their conclusions are not critically dependent on embryo mass, as long as it exceeds $0.2 M_\oplus$ allowing the presence of magma oceans (as we consistently observe in Figure \ref{subfig:Ne_0.2}). By contrast, with the realistic treatment of the cooling of primordial envelopes developed in Section \ref{subsec:gas_accretion}, we find that primordial envelopes atop $M_{\rm rock} \lesssim 0.4 M_\oplus$ embryos would accrete and cool too efficiently to maintain magma oceans at their surface before they reach $M_{\rm env}/M_{\rm rock} \sim 0.2\%$. Consequently, the results of \cite{Young_2023} can only be reconciled with the thermal evolution of primordial envelopes of $\sim 0.5 M_\oplus$ embryos and cannot be consistently translated to lower embryo masses.

This last conclusion further complicates the overall picture of the formation of the Earth. As shown in Figure \ref{fig:comparison_young+23}, the existence of $\sim 0.5 M_\oplus$ embryos accreting and dissolving primordial gas from the solar nebula would drastically overpredict the target concentration expected from the bulk mantle measurements of \cite{Marty_2012}. Thus, one clear implication emerges: if the Earth's water and density deficit were primarily sourced by the dissolution of primordial H, the deep mantle should contain about $\sim 1000$ times more primordial Ne than observed. Conversely, a level of hydrogen dissolution consistent with the neon constraints of \cite{Williams_2019} would be insufficient to account for Earth's water content and a large fraction of the core density deficit, requiring complementary sources of water.

This tension could be resolved by departing from primordial H dissolution as the primary source of the Earth's water, with the early accretion of water-rich solids or a late delivery via ice-rich comets being viable alternatives to the ingassing of a nebular gas envelope \citep{Genda_2016}. 
In fact, \cite{Wu_2018} argued that the combined constraints of the water content of the bulk silicate Earth and its isotopic H composition require at most $\sim 5 \%$ of the planet's water to originate from the ingassing of primordial H in the mantle of embryos, supplementing a late veneer of carbonaceous chondrites (see their Tables 1 and 2). We note that the results of \cite{Wu_2018} complementarily reproduce the Ne terrestrial inventory, assuming an envelope-mantle surface pressure of $1 \ {\rm bar} = {\rm 0.1 \ MPa}$. This pressure value is retrieved in our calculation at the depth of the $M_{\rm env} \approx 10^{-5} M_{\rm rock}$ gas envelope of the favored $M_{\rm rock} = 0.3 M_\oplus$ embryos (see the middle panel of Figure \ref{fig:example_envelope}), which corresponds to the solidification of the mantle ($T_0<T_{\rm melt} \approx 1800{\rm K}$) and thus the end of Ne delivery to the interior. While it is possible that further envelope mass was accreted \citep[such as the value of $M_{\rm env}/M_{\rm rock} =0.2\%$ required by the model of ][]{Young_2023}, we conclude from the results of Section \ref{sec:results} that no further ingassing of the primary envelope in the interior would be allowed once the mantle solidifies at $(P_0,T_0,M_{\rm env}/M_{\rm rock}) \approx (0.1 {\rm MPa}, 1800 {\rm K}, 10^{-5})$.  In light of this observation, our model anchored to the evidence of nebular Ne in the bulk silicate Earth conversely supports minimal contribution from a primary atmosphere to the Earth's water budget and provides a robust confirmation of the conclusions of \cite{Wu_2018}.

We close this section with a brief review of alternative explanations for the core density and high oxidation state of the Earth.
\cite{Wu_2018} find that the H mass dissolved in the core would only reduce the density of the core by $\lesssim 0.27\%$, a fraction of the total density deficit.
This conclusion does not in itself challenge the current understanding of the composition of the core, as its main volatile components remain unknown \citep{Hirose_2021} and therefore allow a wide range of possible delivery mechanisms. For example, \citet[][see their Table 7]{McDonough_2014} argue that compositional models with S-Si or S-O mixtures as the predominant volatiles in the core are the most promising (as opposed to C-rich and H-rich models), to explain the density deficit in light of the current core formation and cosmochemical constraints.
Meanwhile, the high oxidation state of the upper mantle could alternatively be explained by a late veneer of FeO-rich impactors \citep{Waenke_1988}, or by magnesium silicate perovskite (MgSiO$_3$) in the lower mantle, acting as an ``oxygen pump" injecting ferric iron in the mantle and oxidizing infalling material \citep{Wade_2005}.

\section{Conclusion}
\label{sec:conclusion}

Based on the evidence supporting primordial remnants of the solar nebula preserved in the Earth's deep mantle, we have presented a gas accretion and dissolution calculation which models the thermal evolution of the primordial gas envelopes of early Earth embryos, linking their formation environment to the planet's deep mantle $^{22}$Ne reservoir.
Based on our results, the most consistent formation scenario of the Earth to explain a nebular reservoir of Ne in the deep mantle proceeds as follows:

\begin{enumerate}
    \item Earth embryos of mass $\sim 0.3-0.4 M_\oplus$ emerge as the primordial solar nebula is dispersing, at a time when nebular gas density is depleted from the initial MMSN value by a factor of 0.01\% to 10\%. Each of these embryos accretes a thin primordial gas envelope from the gas disk.
    \item The existence of molten magma allows the delivery of primordial volatiles to the rocky interior of embryos. The dissolution of primordial gas is stopped by the rapid solidification of the mantle, after $10^{-5} {\rm \ yr} \lesssim t_{\rm acc}\lesssim 10^{-2} {\rm \  yr}$ if gas accretion occurs in a dust-free regime or after $10^{-2} {\rm \ yr} \lesssim t_{\rm acc}\lesssim 10 {\rm \ yr}$ in the presence of dust grains. 
    Later melting events happen post-dispersal of the solar nebula and remnants of this primordial volatile reservoir explain the nebular Ne signature measured in the deep mantle.
    \item Starting from a set of three $\sim 0.3-0.4 M_\oplus$ embryos, the final assembly of the Earth post-disk dispersal occurs in a series of two mergers, consistent with current models of the Earth's growth through giant impacts.
    \item The deep mantle reservoirs of He and Ne can be attributed to the dissolution of nebular gas in the rocky interiors of embryos, but the non-radiogenic components of Ar, Kr and Xe require an additional source, most likely the inclusion of chondritic volatiles early in the solid assembly of the embryos. Primitive reservoirs of heavy noble gases therefore probe Earth's solid accretion history, while light noble gases are direct probes of primordial gas accretion.
    \item The ingassing of a nebular atmosphere would deliver H to the interior of embryos and produce water through its interaction with mantle silicates. However, the amount accreted would not be sufficient to explain the Earth's water and core density deficit self-consistently with the primordial Ne content of the deep mantle.
    
\end{enumerate}

Our conclusions reflect the best constraints achievable with the present data, though they remain subject to uncertainties stemming from the limited precision of volatile measurements in the Earth's deep mantle.
Future studies of mantle-envelope exchanges of primordial volatiles in early Earth embryos would benefit from more detailed considerations of mantle thermodynamics, but such considerations lie beyond the scope of this work. In particular, a more careful treatment of the mixing of dissolved volatiles to the deep interior is warranted.

\begin{acknowledgments}
We thank the anonymous reviewers for their insights and suggestions, which helped improve the clarity and quality of the manuscript.
EJL thanks Curtis Williams whose seminar at McGill and follow-up discussions helped spark the project.
Huiyi (Cheryl) Wang carried out some of the initial calculations, for which Nicolas Cowan and William Minarik provided feedback. We thank James Day, Steve Desch, Ruth Murray-Clay, James Owen, and Rita Parai for insightful discussions. Some of these discussions took place at the KITP Edgeplanets program in 2025, supported in part by grant NSF PHY-2309135 to the Kavli Institute for Theoretical Physics (KITP). VS acknowledges support by the Fonds de recherche du Québec - Nature et technologies (FRQNT) Master's Training Scholarships and by the Natural Sciences and Engineering Research Council of Canada (NSERC) Postgraduate Scholarship – Doctoral (PGS D). EJL was supported by NSF Research Grant 2509275, NSERC Discovery Grant RGPIN-2020-07045, DGECR-2020-00230, FRQNT Établissement de la relève professorale FRQ-NT NC-298962, FRQNT/NSERC NOVA Grant FRQ-NT 2023-NOVA-325929, NSERC ALLRP 577027-22, and the William Dawson Scholarship from McGill University. Computations were performed on the Graham, Cedar and Rorqual clusters of the Digital Research Alliance of Canada, along with the Triton Shared Computing Cluster (TSCC) of the San Diego Supercomputer Center. 
\end{acknowledgments}

\begin{contribution}
VS carried out all the calculations and drafted the writing.
EJL conceived the project and supervised the overall calculations and writing.


\end{contribution}


\bibliography{References}{}
\bibliographystyle{aasjournalv7}



\end{document}